\documentclass[%
reprint,
superscriptaddress,
 amsmath,amssymb,
 aps,
prb,
]{revtex4-2}

\usepackage{graphicx}
\usepackage{caption}
\usepackage{subcaption}
\usepackage{comment}
\usepackage{dcolumn}
\usepackage{bm}
\usepackage{physics}
\usepackage{xcolor}
\newcommand{\TKedit}[1]{{\color{blue} #1}}

\usepackage{verbatim}

\usepackage{ulem}

\usepackage{tikz}
\usetikzlibrary{angles,calc,intersections,quotes,arrows.meta}
\usetikzlibrary{ "arrows", "automata", "backgrounds", "calendar", "chains", "matrix", "mindmap", "patterns", "petri", "shadows", "shapes.geometric", "shapes.misc", "spy", "trees"}
\usetikzlibrary{shapes,shadows,arrows,positioning,graphs}
\usepackage{pgfplots}
\pgfplotsset{compat=1.17}
\usetikzlibrary{intersections, pgfplots.fillbetween}

\begin{document}
\preprint{APS/123-QED}

\title{Spin conductance in SNN junctions with non-centrosymmetric superconductors}


\author{T.H. Kokkeler}
\email{tim.kokkeler@dipc.org}
\affiliation{Donostia International Physics Center, 20018, Donostia-San Sebastian, Spain\\
}%
\affiliation{Interfaces and Correlated Electron Systems,
 Faculty of Science and Technology, University of Twente, Enschede, The Netherlands\\
}%

\author{Y. Tanaka}
\affiliation{Deparment of Applied Physics, Nagoya University, 464-8603 Nagoya, Japan\\
}%
\author{A.A. Golubov}
\affiliation{Interfaces and Correlated Electron Systems,
 Faculty of Science and Technology, University of Twente, Enschede, The Netherlands\\}%


\date{\today}

\begin{abstract}
An SNN-junction in which the superconducting potential is a mixture between s-wave and p-wave potentials is investigated using the Usadel equation equipped with Tanaka-Nazarov boundary conditions.
The article provides several ways to distinguish between s + chiral and s + helical p-wave superconductors and a way to determine whether a superconductor has a mixed pair potential. Thus, it is of great importance in the determination of the pair potential of superconductors.
It is shown that the different spin sectors satisfy independent equations and can thus be calculated separately even if the d-vector depends on the direction of momentum. This greatly simplifies the equations to be solved. It was found that a difference in conductance for sectors with opposite spins arises if both an s-wave and a p-wave component is present, even in the absence of a magnetic field. The results are confirmed by calculations in the ballistic regime. It is shown that the spin conductance for s + chiral p-wave and s + helical p-wave junctions is qualitatively similar. A setup containing two SN junctions is shown to give a clear difference between the two types of superconductivity. 
\begin{description}
\item[Usage]
Secondary publications and information retrieval purposes.
\item[Structure]
You may use the \texttt{description} environment to structure your abstract;
use the optional argument of the \verb+\item+ command to give the category of each item. 
\end{description}
\end{abstract}

\maketitle

\begin{figure}
    \centering
    \includegraphics[width = 8.4cm]{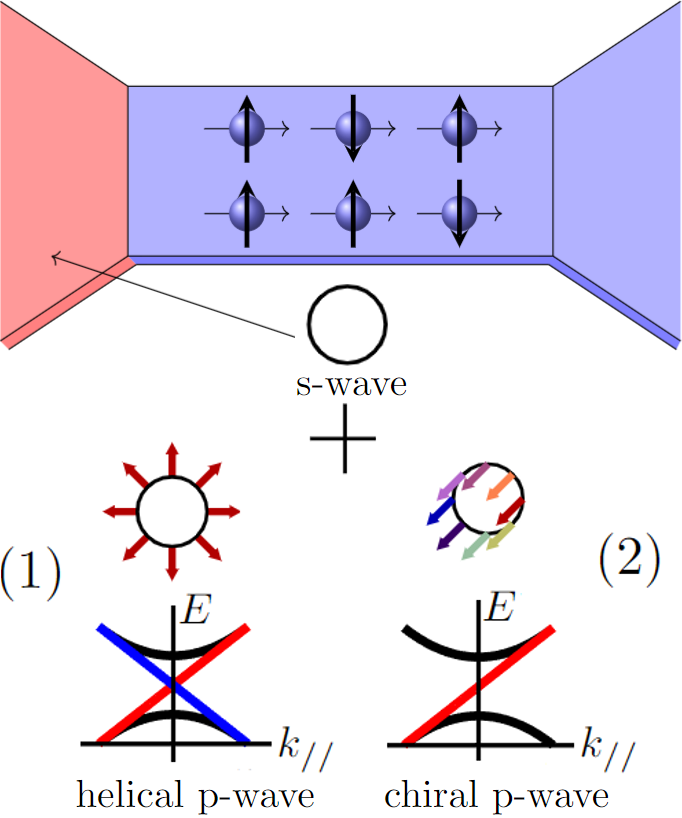}
    \caption{Using an superconductor with a mixed singlet (s-wave) and triplet (p-wave) p-wave pair potential leads to a difference in conductance for opposite spins in the SNN junction. Two types of superconductor pair potential are investigated. First, a mixture of s-wave and helical p-wave (eq. \ref{eq:shelical}) superconductors are investigated. For helical superconductors the magnitude and phase of the d-vector are constant, but the direction of the d-vector is aligned with the direction of momentum. In helical superconductors there are two sets of chiral edge modes, the left and right travelling modes. Second, a mixture of s-wave and chiral p-wave (eq. \ref{eq:schiral}) superconductors is investigated. For chiral superconductors. the d-vector always points out of plane, but has a phase dependent on the direction of momentum, as illustrated by color, are used. The bulk boundary correspondence implies that for this type of superconductors there is only one type of chiral edge modes, only the right moving edge states or only the left moving edge states.}
    \label{fig:illustration}
\end{figure}
The field of superconductivity has attracted a lot of attention since its discovery both experimentally and theoretically \cite{de1964boundary}. Since the discovery of high temperature superconductors \cite{steglich1979superconductivity}, \cite{buchholtz1981identification}, \cite{bednorz1986possible}, and the discovery of novel types of materials such as topological insulators \cite{fu2008superconducting}, attention has shifted more and more towards unconventional superconductors \cite{sigrist1991phenomenological}, \cite{maeno2011evaluation}, \cite{sigrist2005review}, \cite{schnyder2008classification}, \cite{kallin2009sr2ruo4}, such as spin triplet superconductors \cite{eschrig2015spin}, \cite{linder2015superconducting}, whose existence remains under debate even today \cite{maeno2011evaluation}, \cite{suh2020stabilizing}, and odd-frequency superconductivity \cite{bergeret2005odd}, \cite{tanaka2011symmetry}, \cite{linder2019odd}, \cite{cayao2020odd}. Examples of materials that might have a spin triplet superconductivity phase are $\text{Sr}_{2}\text{RuO}_{4}$ \cite{kallin2009sr2ruo4}, the Uranium compound superconductors \cite{huxley2001uge},\cite{jiao2020chiral}, \cite{aoki2011ferromagnetism} and $\text{CrAs}$-based superconductors \cite{luo2019tuning}, \cite{liu2019multiband}, \cite{yang2021spin}.
In order to use these materials its material properties should be known and understood. However, the exact pair potential in the materials above is still unknown and remains an important question to be answered. In this work an experiment is modelled in which different types of pair potential can be distinguished. This experiment can thus be used to narrow the window of possible pair potentials of unconventional superconductors.
A conventional way to describe mesoscopic systems that include superconductors is using the quasiclassical Keldysh Green's function technique \cite{belzig1999quasiclassical}, \cite{chandrasekhar2003introduction}, where the Green's function is given by $G = \begin{bmatrix}G^{R}&G^{K}\\0&G^{A}\end{bmatrix}$, where $G^{R,A,K}$ are the retarded, advanced and Keldysh Green's functions in particle-hole and spin space. The Eilenberger equation \cite{eilenberger1968transformation}, \cite{larkin1969quasiclassical} is used in the clean limit and the Usadel equation \cite{usadel1970generalized} in the dirty limit. In this paper, the Usadel equation will be studied to describe a normal metal proximized \cite{josephson1962possible}, \cite{josephson1964coupled}, \cite{josephson1965supercurrents}, \cite{larkin1977non}, \cite{kastalsky1991observation}, \cite{volkov1993proximity}, \cite{volkov1994proximity}, \cite{van1992excess}, \cite{nazarov1994circuit}, \cite{yip1995conductance}, \cite{schmidt1997physics}, \cite{tanaka2004anomalous} by a superconductor whose pair potential has both an s-wave and a p-wave component, studied before in different geometries and limits \cite{eschrig2010theoretical}, \cite{gentile2011odd}, \cite{annunziata2012proximity}, \cite{rahnavard2014magnetic} \cite{mishra2021effects}. The geometry to be studied is a so-called SNN junction, where the normal metal bar to be described is attached to a superconducting and a normal metal electrode.  Following the notation in \cite{mishra2021effects}, the superconductor pair potential is written as $\Delta = \frac{1}{\sqrt{r^{2}+1}}\Delta_{s}+\frac{r}{\sqrt{r^{2}+1}}\Delta_{p},$
where $r$ denotes the relative strength of the s-wave and p-wave contributions to the potential, $\Delta_{s}$ is the s-wave superconductor pair potential, and $\Delta_{p}$ is one of the p-wave superconductor pair potentials. It is assumed that the contact with the normal metal electrode is very good. Therefore continuity at this interface can be assumed. At the interface between the bar and the superconducting electrode, the Usadel equation is equipped with the Tanaka Nazarov boundary conditions \cite{tanaka2003circuit}. The Tanaka Nazarov boundary conditions are the generalisation of Nazarov's circuit theory \cite{nazarov1999novel} to non-isotropic superconductors, using that odd-parity even-frequency triplet superconductors induce even-parity odd-frequency triplet pairs in a dirty normal metal \cite{tanaka2007theory}. Using these boundary conditions, several types of systems have been studied \cite{tanaka2004theory}, \cite{tanaka2005theory}, \cite{asano2007conductance}, \cite{yokoyama2006resonant}, \cite{sawa2007quasiclassical}, \cite{suzuki2019effects}, \cite{suzuki2020quasiparticle}, \cite{kokkeler2021usadel}. Mixed potentials in an SNN junction were studied in the one-dimensional case using a new form \cite{tanaka2021phys} of the Tanaka Nazarov boundary conditions.\\
However, in the one-dimensional case one can not distinguish between different types of s + p - wave potentials. Therefore, in this work the method is generalised to the two dimensional s + helical p-wave junction. Moreover, the expression for the resistance of the junction following from the Keldysh equations \cite{tanaka2021phys}, has been generalised, allowing the distribution functions to be opposite for opposite spins.
It is shown that the density of states, the charge conductance and the spin conductance can all be used to distinguish between s + chiral and s + helical p-wave superconductors. Next to this, it is shown that the spin conductance can be used to determine whether a superconductor has a mixed pair potential.
With this, our article contributes to the determination of pair potentials in and the understanding of novel unconventional superconductors.

In \cite{tanaka2021phys} a new form of the Tanaka-Nazarov boundary condition was derived, and this new form was used to study one-dimensional SNN junctions. Next to the one-dimensional bar, also a two-dimensional bar in which the Green's function has no y-dependence can be considered, that is, either a bar that is thin enough that the one-dimensional Usadel equation can be studied or junction that is wide, so that boundaries are far away. In this case, the Usadel equation does not change, but multiple modes contribute to the Tanaka-Nazarov conditions \cite{nazarov1999novel} \cite{tanaka2003circuit}, \cite{tanaka2004theory}. The modes are integrated as specified in \cite{tanaka2004theory}. For pairs with opposite spin, the s-wave potential changes sign, whereas the p-wave potential does not. The same parameters in the Tanaka-Nazarov boundary condition as in \cite{tanaka2021phys} were used, that is, the ratio $\gamma_{B}$ between boundary resistance and bar resistance is set to $\gamma_{B} = 2$, the transparency is given by $T = \frac{\cos^{2}{\phi}}{\cos^{2}{\phi}+z^{2}}$ with $z = 0.75$ and the Thouless energy was set to $E_{Th} = 0.02\Delta$. \\
The density of states at the normal metal side of the superconductor-normal metal interface and the conductance of the junction were calculated for various $r$. The calculation for the s + chiral p-wave junction, for which the d-vector is $(0,0,e^{i\phi})$ is relatively similar to the calculation of the one-dimensional s + p-wave junction, though the phase in the diffusive normal metal must be computed numerically, as discussed in the supplementary material. The calculation of the Green's function in the s + helical p-wave junction is more complicated than for the one-dimensional case studied in \cite{tanaka2021phys} or the two-dimensional s + $\text{p}_{\text{x}}$-wave and s + chiral p-wave junctions, because the d-vector has directional dependence.
For mixed s-wave and helical p-wave superconductors
\begin{align}\label{eq:shelical}
    \Bar{\Delta} = \Delta_{s}\mathbf{1}+\Delta_{p}(\cos{\phi}\sigma_{x}+\sin\phi\sigma_{y}).\\
    \Delta_{\pm} = \Delta_{s}\pm\Delta_{p},\nonumber
\end{align}
whereas for mixed s-wave and chiral p-wave superconductors
\begin{align}\label{eq:schiral}
    \Bar{\Delta} = \Delta_{s}\mathbf{1}+\Delta_{p}e^{i\phi}\sigma_{z}\\
    \Delta_{\pm} = \Delta_{s}\pm\Delta_{p}e^{\pm i\phi},\nonumber
\end{align}
Therefore, the retarded Green's function is a four by four matrix. The Usadel equation for four by four matrices is much less convenient to solve than the Usadel equation for two by two matrices, because satisfying the normalisation condition is more complex. Therefore, a way to find two separated equations for two by two matrices, as for the previous cases, was sought.
From expression \ref{eq:shelical} it is found that the average potential over $(-\frac{\pi}{2},\frac{\pi}{2})$ is $\Delta_{s}+\frac{2}{\pi}\Delta_{p}\sigma_{x}$. Thus, it is convenient to transform to the basis of spins polarised in the x-direction via  $G\xrightarrow[]{}UGU$, where
\begin{equation}
    U = \frac{1}{\sqrt{2}}
    \begin{bmatrix}\sigma_{x}+\sigma_{z}&0\\0&\sigma_{x}+\sigma_{z}\end{bmatrix}
\end{equation}
Since $U\tau_{3}U = \tau_{3}$ and $U^{2}$ is the identity, this new Green's function satisfies the Usadel equation with Tanaka Nazarov boundary conditions if also $G_{S}$ is transformed to $UG_{S}U$. It will be shown that after this transformation, $G$ is of the form
\begin{equation}
    G = \begin{bmatrix}\cosh{\theta_{\uparrow}}&0&0&\sinh{\theta_{\uparrow}}e^{i\chi_{\uparrow}}\\0&\cosh{\theta_{\downarrow}}&\sinh{\theta_{\downarrow}}e^{i\chi_{\downarrow}}&0\\0&-\sinh{\theta_{\downarrow}e^{-i\chi_{\downarrow}}}&-\cosh{\theta_{\downarrow}}&0\\-\sinh{\theta_{\uparrow}}e^{-i\chi_{\uparrow}}&0&0&-\cosh{\theta_{\uparrow}}\end{bmatrix},
\end{equation}
that is, $YGY = G$, with $Y = \text{diag}(1,-1,-1,1)$.\\ Because for $A,B$ satisfying $YAY = A$ and $YBY = B$ it also holds that $YABY = AB$, the Usadel equation itself does not violate this symmetry, the only point of concern is the boundary condition at the interface between the normal metal and the superconductor, that is, it has to be shown that if $YGY = G$ in the diffusive normal metal, then also $Y\langle S\rangle Y = \langle S\rangle$, where $\langle S\rangle $ is the retarded part of the
Tanaka-Nazarov boundary conditions discussed in supplementary material and the notation $\langle\rangle$ is used to denote angular averaging. Note that it is not necessary to satisfy $YS(\phi)Y = S(\phi)$ for every angle, since only the angular averaged term appears in the boundary condition. In section \ref{sec:Decoupling} it is shown that in fact $YS(\phi)Y = S(-\phi)$, so that $Y\langle S\rangle Y = \langle S\rangle$ is indeed satisfied.\\
The discussion so far shows that the Green's function in the normal metal satisfies $YGY = G$, and therefore, the $\theta$-parametrisation can be used for the different blocks for the up and down orientation. However, in principle for the computation of the boundary condition both are needed, that is, the equations are coupled at the boundary. However, in section \ref{sec:Qsymm} it is shown that also the symmetry $QGQ = G$, where $Q =\begin{bmatrix}0&\sigma_{x}\\-\sigma_{x}&0\end{bmatrix}$. This relates the solution for the different components by $\theta_{\downarrow} = -\theta_{\uparrow}$, $\chi_{\downarrow} = -\chi_{\uparrow}$. This implies that only one of the components needs to be calculated.

\begin{figure}
    \centering
    \includegraphics[width = 8.4cm]{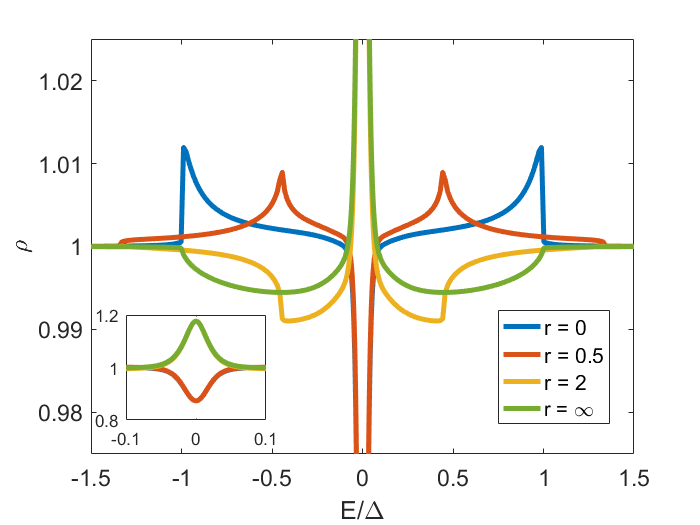}
    \caption{The density of states in the s + helical p-wave junction for different values of $r = \frac{\Delta_{p}}{\Delta_{s}}$\TKedit{ as a function of enegy $E$}. In the inset it is shown that there is a zero energy peak in the density of states if the p-wave component is dominant, while there is a dip if the s-wave component is dominant, similar to the one-dimensional case \cite{tanaka2021phys}. Both the peak and the dip have a width on the order of the Thouless energy.  For nonzero energies the density of states becomes dependent on the actual value of $r$. }
    \label{fig:RhoHelical}
\end{figure}

\begin{figure}
    \centering
\subcaptionbox{S + chiral p-wave.
  }[0.45\linewidth]
  {\hspace*{-2em}\includegraphics[width =8.4cm]{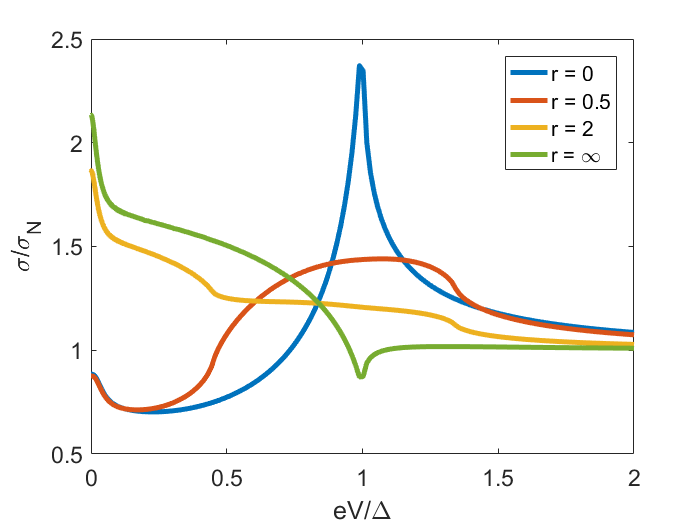}}
\hfill
\subcaptionbox{S + helical p-wave.
  }[0.45\linewidth]
  {\hspace*{-2em}\includegraphics[width =8.4cm]{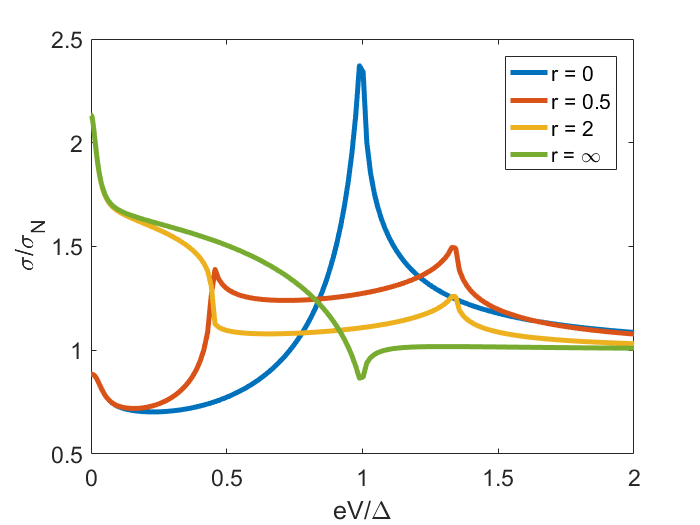}}
    \caption{The conductance of the s + chiral and s + helical p-wave junction for different values of the parameter $r = \frac{\Delta_{p}}{\Delta_{s}}$. There is a clear zero bias conductance peak if the p-wave component is dominant, consisting of both a contribution with a width of the order of the Thouless energy, and a broad peak inherent to the dominant p-wave pair potential. If the s-wave component is dominant there is only a small peak of the width of the Thouless energy \cite{volkov1993proximity}. For the s + helical p-wave superconductor the zero bias conductance peak is robust against inclusion of an s-wave component of the pair potential, for the s + chiral p-wave superconductor it is not. Moreover, for $eV\in(\Delta_{-},\Delta_{+})$, there is a smooth peak for s + chiral p-wave junctions and two sharper peaks for s + helical p-wave junctions.}
    \label{fig:HelicalR}
\end{figure}

\begin{figure}
    \centering
\subcaptionbox{S + chiral p-wave.
  }[0.45\linewidth]
  {\hspace*{-2em}\includegraphics[width =8.4cm]{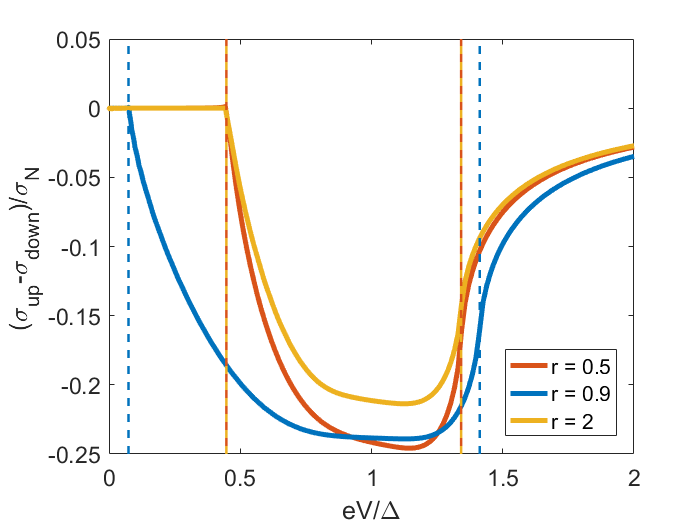}}
\hfill
\subcaptionbox{S + helical p-wave.
  }[0.45\linewidth]
  {\hspace*{-2em}\includegraphics[width =8.4cm]{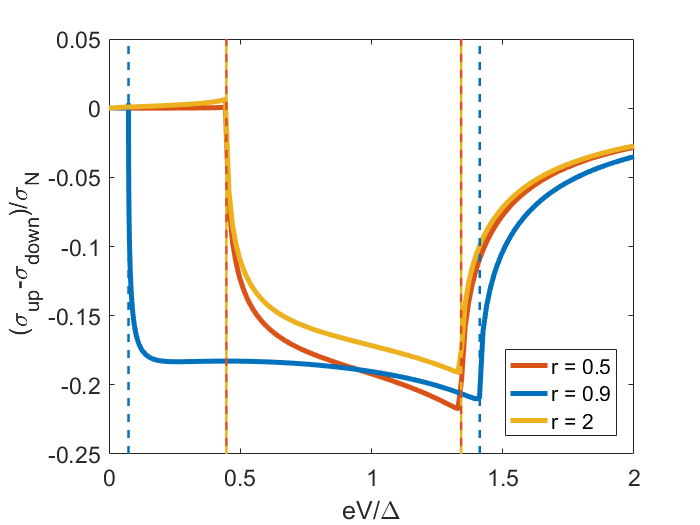}}
    \caption{The difference in spin conductance for spin sectors up $(\sigma_{\text{up}})$ and down $\sigma_{\text{down}}$ for s + chiral p-wave junctions and s + helical. For spin-singlet or spin-triplet superconductors without parity mixing the spin conductance is spin independent, but for mixed potential superconductors there is a difference in the conductance for different spin orientations. The difference in spin conductance is very small for $eV<|\Delta_{-}|$, indicated with the left dotted lines, and decays to zero for $eV>\Delta_{+}$ indicated with the right dotted lines. The results for s + chiral p-wave and s + helical p-wave are qualitatively similar, though for s + helical p-wave junctions the transition between the three regimes is much sharper.}
    \label{fig:Spinconductance}
\end{figure}

The resulting local density of states normalised to the normal metal local density of states is shown in figure \ref{fig:RhoHelical}, denoted by $\rho$. If the s-wave component of the pair potential is dominant, that is, if $r<1$, there is a dip in the density of states at zero energy , whereas $r>1$ results in a peak in the density of states at zero energy. The dip is not at $\rho = 0$ because of the presence of the normal metal electrode. The zero energy peak or dip is equally high for mixed potentials as it is for pure s-wave or helical p-wave superconductors, similar to the one-dimensional case \cite{tanaka2021phys}. This can be understood from equations \ref{eq:Deltaplus} and \ref{eq:Deltamin}. Contrary to the chiral p-wave junctions, the potentials $\Delta_{\pm}$ are independent of angle for helical p-wave junctions. This independence arises because the eigenvalues of $(\cos{\phi}\sigma_{x}+\sin{\phi}\sigma_{y})$ are always 1 and -1, regardless of $\phi$. Therefore, $\frac{\Delta_{+}}{|\Delta_{+}|}$ is 1 and $\frac{\Delta_{-}}{|\Delta_{-}|}$ is either 1 (if $r<1$) or -1 (if $r>1$) regardless of the angle $\phi$. This should be contrasted with the chiral p-wave case, where $\Delta_{\pm} = \frac{1}{\sqrt{r^{2}+1}}\left(1\pm r e^{i\phi}\right)$, which means that $\frac{\Delta_{+}}{|\Delta_{+}|}$ and $\frac{\Delta_{-}}{|\Delta_{-}|}$ will dependent on $r,\phi$. Since the zero energy results only depend on $\frac{\Delta_{+}}{|\Delta_{+}|}$ and $\frac{\Delta_{-}}{|\Delta_{-}|}$, as discussed in supplementary material \ref{sec: Very low energy}, there is a binary distinction at $E = 0$ between $r<1$ and $r>1$ for s + helical p-wave superconductors, and a non-binary distinction for s + chiral p-wave junctions.
For nonzero energies the density of states becomes dependent on the actual value of $r$. For $r<1$ there is a peak at $E\approx \Delta_{-}$, then slowly decays for $E\in(\Delta_{-},\Delta_{+})$, and a very sharp decay towards the normal states for $E>\Delta_{+}$ follows. If $r>1$ there is a dip around $E\approx|\Delta_{-}|$, after which the density of states increases towards the normal states value.
Also the conductance was calculated for the s + helical p-wave superconductor. Observe that if the retarded part $C^{R}$ satisfies $C^{R}(-\phi) = YC^{R}(\phi)Y$, then the advanced part $C^{A}$ and Keldysh part $C^{K}$ satisfy $YC^{A}(-\phi)Y = C^{A}(\phi)$ and $YC^{K}(-\phi)Y = C^{K}(\phi)$ as well.
Thus, also the equations for the Keldysh part splits into two systems of equations for two by two matrices that are coupled at the boundary. Even more, the equations $\theta_{\uparrow} = -\theta_{\downarrow}$ and $\chi_{\uparrow} = -\chi_{\downarrow}$ are still satisfied by the symmetry discussed in section \ref{sec:Qsymm}, as in the $\text{p}_{\text{x}}$-wave and chiral p-wave case. Thus, the expression for the resistance of the junction can be derived similar to \cite{tanaka2021phys}, with an appropriate definition of $I^{K}$. The expressions for $I^{K}$ need to be compared to the definition in \cite{tanaka2021phys}. The new definition of $I^{K}$ is discussed in supplementary material \ref{sec:Helical conductance}, where it is also shown that the new expression reduces to the expression in \cite{tanaka2021phys} if there is no helical component in the superconductor pair potential.\\
The results for the conductance $\sigma = \frac{\partial I}{\partial V}$ are shown in figure \ref{fig:HelicalR} normalised to the conductance $\sigma_{N}$ of the normal metal if it is not proximised by the superconductor. The conductance in the s + helical p-wave junction is very similar to the conductance in the s + chiral p-wave junction, if $\Delta_{p}>\Delta_{s}$ there is both a sharp peak and a broader peak. This is consistent with the appearance of Andreev bound states with dispersion in \cite{burset2014transport}. An important difference with the s + chiral p-wave however, is that for s + helical p-wave junctions the zero bias conductance is quantised, that is, it assumes a constant value for $r<1$ and a constant value for $r>1$. Thus, where the addition of an s-wave component lowered the zero bias conductance peak both when mixed with a $\text{p}_{\text{x}}$-wave and when mixed with a chiral p-wave superconductor, in helical superconductor junctions the addition of an s-wave component does not influence the height of the zero bias conductance peak. Moreover, the double peak structure, as observed in \cite{chiu2021observation} and discussed for pure chiral p-wave superconductors in \cite{tanaka2005theory},  is robust against the inclusion of an s-wave component. This is in correspondence with the results for the density of states and the pair amplitude.
\onecolumngrid\
\twocolumngrid\
The derivation of the conductance in supplementary material \ref{sec:NEWconductanceformula} explicitly allows for a difference in conductance for different spin orientations. This feature has never appeared in superconductors that are spin-singlet or spin-triplet without parity mixing, so it is expected that this difference should vanish for these types of superconductors. However, for mixed type superconductors it is different. The Green's function both has a singlet and a triplet component, that is, the spin structure is of the form $a+b\sigma$ with $ab\neq 0$. This means that for each angle there is a spin polarisation in the Green's function. Averaged over the full angle this polarisation will disappear. However, for the boundary condition, the average is only taken over $(-\frac{\pi}{2},\frac{\pi}{2})$, and it is not clear whether this term vanishes.\\
An exact expression for the spin conductance can be found in supplementary material \ref{sec:NEWconductanceformula}. The spin conductance $\sigma_{\text{up}}-\sigma_{\text{down}}$ is shown for the quasi-one dimensional s+$\text{p}_{\text{x}}$-wave junction in figure \ref{fig:Spinconductance} for different values of $r$. It is confirmed that if the superconductor pair potential is of a pure s-wave or p-wave type the difference in conductance for opposite spins is zero, as expected. However, if $r\in(0,\infty)$, there is a finite difference in conductance for different spins. For low voltages this difference is very small, but between $eV = |\Delta_{-}|$ and $eV = \Delta_{+}$ the spin conductance is not small. The spin conductance for s + chiral p-wave superconducting junctions is qualitatively similar as that for s + helical p-wave superconducting junctions, though for the s + helical p-wave case the change around $eV = \Delta_{+}$ is much sharper.
\onecolumngrid\
\twocolumngrid\
To investigate the difference in conductance for different spins further, the ballistic metal case was considered as well. A previous calculation on ballistic junctions including non-centrosymmetric superconductors showed that a difference in conductance appears for spin up and down in the z-direction for modes in a specific direction, but that angular averaging removes this difference \cite{tanaka2009theory}. However, it will be shown that this is not contradictory with our findings, as we will show that the same model does predict a difference in the conductance of spins in the $y$-direction.\\
The superconductor pair potential used in \cite{tanaka2009theory} is given by
\begin{equation}
    \Delta = \Delta_{s}\mathbf{1}+\Delta_{p}(\sin\phi\sigma_{x}-\cos\phi\sigma_{y}).
\end{equation}
In analogy with the s + helical p-wave junction described before in the dirty limit, one should expect that there is a spin current with only a $\sigma_{y}$-component. Therefore, the calculations in \cite{tanaka2009theory} were repeated, but now also the $\sigma_{x}$ and $\sigma_{y}$-components were calculated. 
A discussion of the calculations can be found in the supplementary materials.

\begin{figure}
    \centering
    \includegraphics[width = 8.4cm]{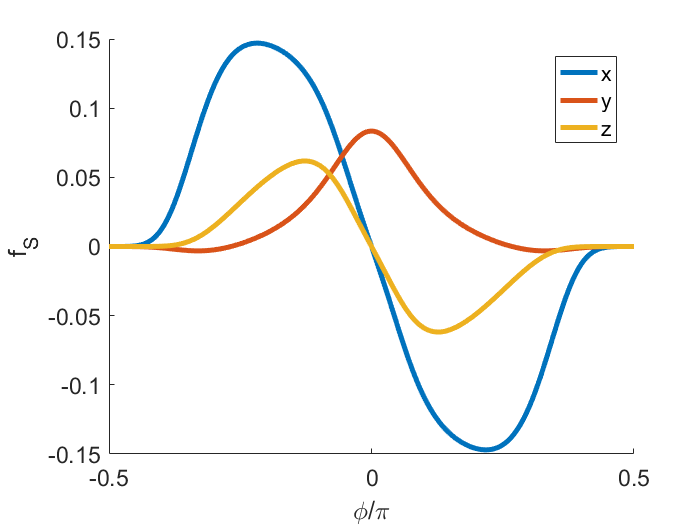}
    \caption{The three different components of the angular dependent spin conductance $f_{S}$ normalised to the normal state conductance as a function of direction for the ballistic s + helical p-wave junction. The x and z components of the spin currents are odd with respect to angle made with the interface and thus average out, the y component however is even with respect to the angle made with the interface.}
    \label{fig:BallisticExample}
\end{figure}

\begin{figure}
    \begin{tikzpicture}[thick, scale=0.9]
    \fill [red!10] (3,1) rectangle (5,2);
    \fill [red!10] (2,5) rectangle (1,3);
    \fill[red!10] (0,5) rectangle (3,8);
    \fill [blue!10] (0,0) rectangle (3,3);
    \fill[red!10] (5,0) rectangle (8,3);
    \draw [ultra thick] (0,0) rectangle (3,3);
    \draw [ultra thick] (3,1) rectangle (5,2);
    \draw [ultra thick] (1,3) rectangle (2,5);
    \draw [ultra thick] (5,0) rectangle (8,3);
    \draw [ultra thick] (0,5) rectangle (3,8);
    \node at (1.5,1.5) {S + Chiral p};
    \node at (6.5,1.5) {Normal metal};
    \node at (1.5,6.5) {Normal metal};
    \draw[->] (3.5,0.5)--(4.5,0.5);
    \draw[->] (0.5,3.5)--(0.5,4.5);
    \node at (4,1.5) {$\text{Tr}I\sigma_{z}\neq 0$};
    \node[rotate = 90] at (1.5,4) {$\text{Tr}I\sigma_{z}\neq 0$};
    \draw[->] (-0.5,0)--(-0.5,3);
    \draw[->] (-0.5,3)--(-0.5,0);
    \node at (-1,1.5) {$W$};
    \fill[red!10] (10,5) rectangle (13,8);
    \fill [green!10] (10,0) rectangle (13,3);
    \fill [red!10] (13,1) rectangle (15,2);
    \fill [red!10] (12,5) rectangle (11,3);
    \fill[red!10] (15,0) rectangle (18,3);
    \draw [ultra thick] (10,0) rectangle (13,3);
    \draw [ultra thick] (13,1) rectangle (15,2);
    \draw [ultra thick] (11,3) rectangle (12,5);
    \draw [ultra thick] (15,0) rectangle (18,3);
    \draw [ultra thick] (10,5) rectangle (13,8);
    \node at (11.5,1.5) {S + Helical p};
    \node at (16.5,1.5) {Normal metal};
    \node at (11.5,6.5) {Normal metal};
    \node at (14,1.5) {$\text{Tr}I\sigma_{x}\neq 0$};
    \node[rotate = 90] at (11.5,4) {$\text{Tr}I\sigma_{y}\neq 0$};
    \draw[->] (13.5,0.5)--(14.5,0.5);
    \draw[->] (10.5,3.5)--(10.5,4.5);
    \end{tikzpicture}
    \caption{Schematic of an experiment that can be used to distinguish between s + chiral and s + helical p-wave junctions. For the s + chiral p-wave junction the spin polarisation of the current is the same in both bars, whereas for the s + helical p-wave junction the polarisation of the current has a perpendicular direction in one bar compared to the other.}
    \label{fig:Setup2}
\end{figure}
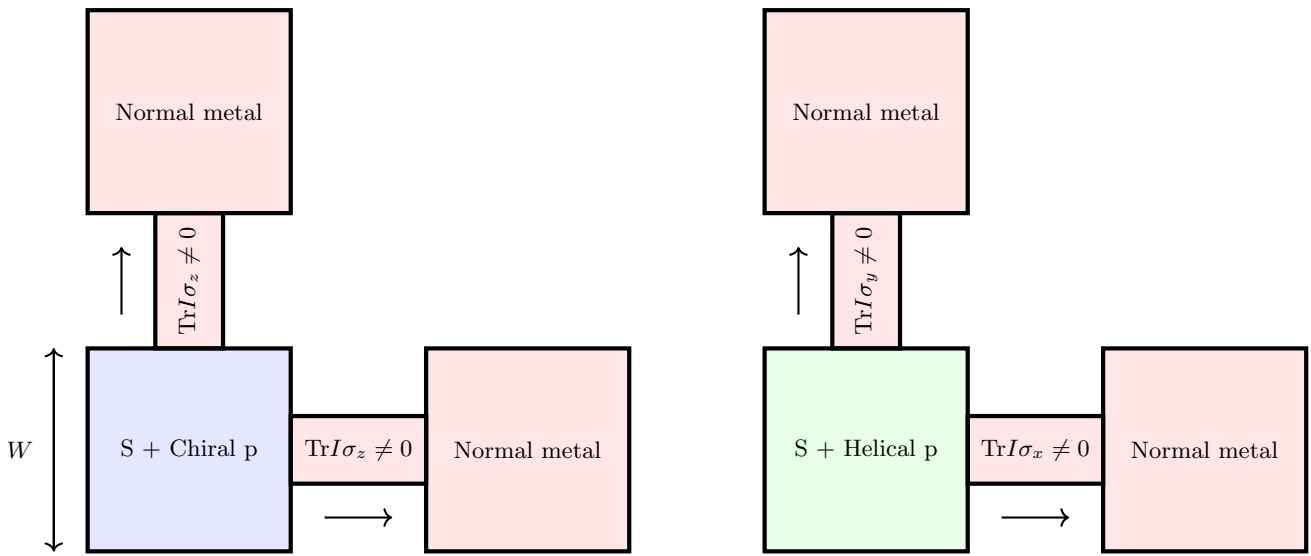

The formalism has been tested in a configuration similar to \cite{tanaka2009theory}, albeit without a spin orbit coupling apart from that in the superconductor pair potential, so that any spin effects are purely due to the superconductor pair potential, as in the dirty case studied in this work. An example result is shown in figure \ref{fig:BallisticExample}, where the angular dependent conductance $f_{S}(\phi) = \frac{\partial I(\phi)}{\partial V}$ is shown in terms of the normal state conductance. For other parameters the conductance changes, but the symmetries still exist, and the y-component is the only nonzero component of the spin current, as expected. Thus, the calculations in the ballistic limit support the findings in the dirty limit.
The spin conductance for s + chiral p-wave superconductors and s + helical p-wave superconductors are qualitatively similar.
For s + helical p-wave superconductors, the change at $eV\approx \Delta_{+}$ is much sharper than for s + chiral p-wave superconductors. However, this difference might not be clearly visible in experiment due to uncertainties. However, a clear difference between s + chiral p-wave superconductors and s + helical p-wave superconductors can be observed in the setup in figure \ref{fig:Setup2}. Instead of using a single SNN junction, a corner junction is studied. It is assumed that the size $W$ of the superconductor is much larger than the coherence length, so that the different arms of the junction do not influence each other. Now, for the s + chiral p-wave superconductor the $d$-vector points in the $z$-direction independent of the direction of momentum, so both bars will show a difference in conductance for spins polarised in the z-direction. For the s + helical p-wave case however, the $d$-vector is dependent on the direction of momentum. In the previous sections it was shown that the spin polarisation depends on the average over the Green's function over modes that make an angle less than $\frac{\pi}{2}$ with the outward normal. This means that for one of the bars the current will be partly polarised along the x-direction, whereas for the other bar the current will be partly polarised along the y-direction. Thus, whereas for the s + chiral p-wave superconducting junction the spin polarisation is in the same direction, for the s + helical p-wave junction the spin polarisation is in perpendicular directions. This means that a clear difference between s + chiral p-wave and s + helical p-wave superconductors can be observed.
\onecolumngrid\
\twocolumngrid\
In this paper it was shown that the SNN junction can be used to distinguish mixed s + p - wave superconductors from the pure s-wave or p-wave superconductors. Moreover, it is shown that a spin conductance can be used to distinguish different types of s + p - wave potentials, such as s + chiral p and s + helical p. The new form of the Tanaka-Nazarov boundary conditions was applied to two-dimensional helical s+p-wave junctions. It was shown that the equations for the helical p-wave junctions can be solved using the two by two matrix formalism, even though there is a strong coupling between the $d$-vector and the direction of the momentum in the superconductor. This allows for the usage of the $\theta$-parametrisation. Moreover, a new expression for the conductance was found, by allowing the distribution functions in the Keldysh component to be spin-dependent. In the case of a pure potential the new expression reduces to the known expression. In the case of a mixed potential the correction to the total conductance is only small, however, it was found that there is a difference in the conductance for different spin orientations in case both an s-wave and a p-wave component are present in the pair potential. It was shown that these results are consistent with ballistic junctions. This is an important result, as it shows a way of obtaining spin currents without any applied magnetic field. For future work, it is important to investigate the supercurrent in junctions containing two superconductors including s + helical p-wave superconductivity, and to calculate whether dissipationless spin currents are possible. Also, the addition of a magnetic field to the setup would be an interesting road to follow.
\section{Acknowledgements}
This work
was supported by Grants-in-Aid from JSPS for Scientific
Research (A) (KAKENHI Grant No. JP20H00131), Scientific
Research (B) (KAKENHI Grants No. JP18H01176
and No. JP20H01857), and the JSPS Core-to-
Core program “Oxide Superspin” international network.
T.K. acknowledges financial support
from Spanish AEI through project PID2020-114252GB-I00

(SPIRIT).

\newpage
\bibliography{apssamp}
\newpage
\appendix
\onecolumngrid\
\section{Superconductivity and spin}\label{sec:Superconductivityandspin}
In BCS theory \cite{BCS1957a}, \cite{BCS1957b}, a Hamiltonian for four-component spinors is used. Thus, the Keldysh Green's function for this Hamiltonian is an eight by eight matrix, with the retarded, advanced and Keldysh components all being four by four matrices. However, in many works \cite{belzig1999quasiclassical}, \cite{chandrasekhar2003introduction}, \cite{tanaka2003circuit}, \cite{tanaka2004theory}, a four by four Keldysh Green's function is used. This is possible because different spin sectors can be studied separately in conventional and some types of unconventional superconductors. In this section, the decoupling of the equations for different spin sectors will be revisited. For clarity of presentation, only the retarded part will be discussed, the advanced and Keldysh parts follow analogously.
\\
The BCS theory \cite{Cooper1956} \cite{BCS1957a}, \cite{BCS1957b} describes an attractive interaction between particles with opposite spin, the Hamiltonian $H$ for the spinor $(\psi_{k\uparrow},\psi_{k\downarrow},\psi^{\dagger}_{-k\uparrow},\psi^{\dagger}_{-k\downarrow})^{T}$ is
\begin{equation}
    H = \begin{bmatrix}E_{s}&0&0&\Delta\\0&E_{s}&-\Delta&0\\0&\Delta&E_{s}&0\\-\Delta&0&0&E_{s}\end{bmatrix},
\end{equation}
where $E$ is the sum of the kinetic and potential energy of the particles, and $\Delta$ the strength of the superconducting interactions, the so-called superconductor pair potential. This means that the retarded part of the quasiclassical Green's function in the bulk reads \cite{belzig1999quasiclassical}, \cite{chandrasekhar2003introduction}:
\begin{align}
    G_{S}^{R}(E) = \begin{bmatrix}\frac{E}{\sqrt{E^{2}-|\Delta|^{2}}}&0&0&\frac{\Delta}{\sqrt{E^{2}-|\Delta|^{2}}}\\0&\frac{E}{\sqrt{E^{2}-|\Delta|^{2}}}&-\frac{\Delta}{\sqrt{E^{2}-|\Delta|^{2}}}&0\\0&\frac{\Delta}{\sqrt{E^{2}-|\Delta|^{2}}}&-\frac{E}{\sqrt{E^{2}-|\Delta|^{2}}}&0\\-\frac{\Delta}{\sqrt{E^{2}-|\Delta|^{2}}}&0&0&-\frac{E}{\sqrt{E^{2}-|\Delta|^{2}}}\end{bmatrix}.
\end{align}
There are two important remarks to be made about this Green's function:
\begin{itemize}
    \item The matrix $G_{S}^{R}$ has a special structure, it can be seen as two blocks of two by two matrices. More precise, $G_{S}^{R}$ satisfies
    \begin{align}
        G_{S}^{R} & = YG_{S}^{R}Y\\
        Y& = \begin{bmatrix}1&0&0&0\\0&-1&0&0\\0&0&-1&0\\0&0&0&1\end{bmatrix}\label{eq:Ydefinition}.
    \end{align}
    This allows us to decouple the dynamics of $\psi_{k\uparrow}, \psi^{\dagger}_{-k\downarrow}$ from the dynamics of $\psi_{k\downarrow},\psi^{\dagger}_{-k\uparrow}$. These will be referred to as the spin up (former) and spin down (latter) components respectively. 
    \item $\langle\psi_{k\uparrow}\psi_{-k\downarrow}\rangle = -\langle\psi_{k\downarrow}\psi_{-k\uparrow}\rangle$, that is, the superconductivity is singlet.
\end{itemize}
Now, recall that for p-wave superconductivity, the superconductor pair potential may be written as $\Delta = \Delta i\sigma_{y}(d_{x}\sigma_{x}+d_{y}\sigma_{y}+d_{z}\sigma_{z}) = \Delta\begin{bmatrix}d_{x}+id_{y}&d_{z}\\d_{z}&-d_{x}+id_{y}\end{bmatrix}$, where $\sigma_{x,y,z}$ denote the Pauli matrices. If the direction of the so-called d-vector $d = (d_{x},d_{y},d_{z})$ is angular independent, as for the $p_{x}$-wave (denoted shortly p-wave here) and chiral p-wave superconductors, then the spin basis can be chosen such that $d_{x} = d_{y} =0$, only $d_{z}$ is non-vanishing. In this basis the Hamiltonian can be written as
\begin{equation}
    H = \begin{bmatrix}E_{s}&0&0&d_{z}\Delta\\0&E_{s}&d_{z}\Delta&0\\0&-d_{z}\Delta&E_{s}&0\\-d_{z}\Delta&0&0&E_{s}\end{bmatrix},\\
\end{equation}
and thus, denoting $\Delta_{p} = d_{z}\Delta$
\begin{align}
    G_{S}^{R}(E) = \begin{bmatrix}\frac{E}{\sqrt{E^{2}-|\Delta_{p}|^{2}}}&0&0&\frac{\Delta_{p}}{\sqrt{E^{2}-|\Delta_{p}|^{2}}}\\0&\frac{E}{\sqrt{E^{2}-|\Delta_{p}|^{2}}}&\frac{\Delta_{p}}{\sqrt{E^{2}-|\Delta_{p}|^{2}}}&0\\0&-\frac{\Delta_{p}}{\sqrt{E^{2}-|\Delta_{p}|^{2}}}&-\frac{E}{\sqrt{E^{2}-|\Delta_{p}|^{2}}}&0\\-\frac{\Delta}{\sqrt{E^{2}-|\Delta_{p}|^{2}}}&0&0&-\frac{E}{\sqrt{E^{2}-|\Delta_{p}|^{2}}}\end{bmatrix}.
\end{align}
For this type of superconductors
\begin{itemize}
    \item It still holds that $YG_{S}^{R}Y = G_{S}^{R}$, the two spin sectors are not mixed.
    \item  $\langle\psi_{k\uparrow}\psi^{\dagger}_{-k\downarrow}\rangle = \langle\psi_{k\downarrow}\psi^{\dagger}_{-k\uparrow}\rangle$, that is, the superconductivity is triplet.
\end{itemize}
If now $s$-wave and $p$-wave superconductivity in which $d_{z}$ is the only non-zero component of the d-vector, for example, $d_{z} = \cos{\phi}$ or $d_{z} = e^{2i\phi}$, are mixed, the Green's function becomes
\begin{align}
    G_{S}^{R}(E) &= \begin{bmatrix}\frac{E}{\sqrt{E^{2}-|\Delta_{\uparrow}|^{2}}}&0&0&\frac{\Delta_{\uparrow}}{\sqrt{E^{2}-|\Delta_{\uparrow}|^{2}}}\\0&\frac{E}{\sqrt{E^{2}-|\Delta_{\downarrow}|^{2}}}&\frac{\Delta_{\downarrow}}{\sqrt{E^{2}-|\Delta_{\downarrow}|^{2}}}&0\\0&-\frac{\Delta_{\downarrow}}{\sqrt{E^{2}-|\Delta_{\downarrow}|^{2}}}&-\frac{E}{\sqrt{E^{2}-|\Delta_{\downarrow}|^{2}}}&0\\-\frac{\Delta_{\uparrow}}{\sqrt{E^{2}-|\Delta_{\uparrow}|^{2}}}&0&0&-\frac{E}{\sqrt{E^{2}-|\Delta_{\uparrow}|^{2}}}\end{bmatrix},\\
    \Delta_{\uparrow}&=\Delta_{s}+\Delta_{p}\label{eq:upspinpotential}\\
    \Delta_{\downarrow}&=-\Delta_{s}+\Delta_{p}\label{eq:downspinpotential}.
\end{align}
In this case 
\begin{itemize}
    \item $YG_{S}^{R}Y = G_{S}^{R}$ is still satisfied.
    \item $\Delta_{\uparrow}\neq\Delta_{\downarrow}$ and $\Delta_{\uparrow}\neq-\Delta_{\downarrow}$, so there is both singlet and triplet superconductivity.
\end{itemize}
This allows us to discuss the mixtures between s-wave and either $\text{p}_{\text{x}}$-wave or chiral p-wave superconductor pair potentials in terms of the decoupled equations for the different spin sectors. In these cases the superconductor pair potentials will be treated as scalars, that is, $\Delta_{s} = \Delta_{0}$ and $\Delta_{p} = \Delta_{0}\cos{\phi}$ or $\Delta_{p} = \Delta_{0}e^{i\phi}$. The superconductor pair potentials can be different for the different spin sectors, as can be seen from equations \ref{eq:upspinpotential} and \ref{eq:downspinpotential}. The case of helical p-wave superconductors is less apparent, as the direction of the $d$-vector depends on $\phi$ via $d = (\cos\phi,\sin\phi,0)$. The application of the symmetry discussed in this section for helical p-wave superconductors is discussed in section \ref{sec:Decoupling}.
\section{The SNN junction} \label{sec:SNNBC}
The geometry considered is an SNN junction. For $x<0$ there is a superconductor, the type of pair potential in this superconductor can be varied. For $0<x<L$ there is a dirty normal metal, that is, a metal in which the scattering rate is high \cite{belzig1999quasiclassical}, and for $x>L$ there is a normal metal reservoir, for which it is assumed that the Green's function assumes the bulk normal metal value. The model describes the properties of the dirty normal metal. To describe the Green's function in this bar, the one-dimensional Usadel equation is used \cite{usadel1970generalized}: 
\begin{align}
    &\dv{}{x}\left(G\dv{G}{x}\right)+2i\eta E(\Tilde{\tau}_{3}G-G\tau_{3}) =0,\label{eq:Usadel}\\
    & G^{2} = \mathbf{1},
\end{align}
where $G$ is the Green's function in Nambu Keldysh space, given by $G = \begin{bmatrix}G^{R}&G^{K}\\0&G^{A}\end{bmatrix}$, with $G^{R,A,K}$ being the retarded, advanced and Keldysh components of the Green's function. The parameter $\eta$ is proportional to the energy scale of the pair potential $\Delta_{0}$ and $T_{c}$ is the critical temperature of the superconductor. According to BCS theory it is given by $\eta = \frac{\Delta_{0}}{2\pi T_{c}} = 1.76$. The matrix $\Tilde{\tau}_{3}$ is given by
\begin{equation}
    \Tilde{\tau}_{3} = \begin{bmatrix}\mathbf{1}&0&0&0\\0&-\mathbf{1}&0&0\\0&0&\mathbf{1}&0\\0&0&0&-\mathbf{1}\end{bmatrix},
\end{equation}
where $\mathbf{1}$ is a unit matrix. If the different spin sectors can be considered separately as discussed in section \ref{sec:Superconductivityandspin}, $\mathbf{1}$ can be replaced by 1.
Any length scale will be expressed in units of the superconducting coherence length $\xi = \sqrt{\frac{D}{2\pi T_{c}}}$, whereas any energy scale is expressed in units of $\Delta_{0}$.
This equation is equipped with two boundary conditions.
\begin{itemize}
    \item On the right hand side, there is a normal metal reservoir. It is assumed that the contact between the bar and this reservoir is good, therefore the Green's function is continuous, $G(L) = G_{N}$, where $G_{N} = \tau_{3}$ is the Green's function in the normal metal reservoir.
    \item On the left hand side, the so-called Tanaka-Nazarov boundary conditions \cite{nazarov1999novel}, \cite{tanaka2003circuit}, \cite{tanaka2004theory} are used, because they explicitly allow for unconventional superconductors, that is, superconductors in which the pair potential is not s-wave. In the compact form written by Tanaka \cite{tanaka2021phys} they read in Keldysh Nambu space:
    \begin{align}
        G\nabla G|_{x = 0} &= \frac{1}{\gamma_{B}L}\langle S\rangle,\\
        S(\phi) &= -2T(2(2-T)\mathbf{1}+T(CG+GC))^{-1}(CG-GC),\\
        C &= H_{+}^{-1}(\mathbf{1}-H_{-})\\
        H_{\pm} & = \frac{1}{2}(G_{S}(\phi)\pm G_{S}(\pi-\phi))
    \end{align}
    Here $\mathbf{1}$ is the identity matrix, $\phi$ is the angle made with the normal of the interface, the notation $\langle\rangle$ denotes an angular average over the angle $\phi\in (-\frac{\pi}{2},\frac{\pi}{2})$, $G_{S}$ is the Green's function in the superconductor and $T$ is the transparency of the interface. 
\end{itemize}
In principle, the Usadel equation described here is an equation for matrices that are eight by eight matrices. However, in section \ref{sec:Superconductivityandspin} it was shown that for s-wave and for $\text{p}_{\text{x}}$-wave and chiral p-wave superconductors with a d-vector pointing in the z-direction, the retarderd, advanced and Keldysh Green's function satisfy $YG^{R,A,K}Y = G^{R,A,K}$, with $Y$ defined as in \ref{eq:Ydefinition}. Moreover, also $Y\tau_{3}Y = \tau_{3}$ and $YABY = AB$ if $YAY = A$ and $YBY = B$. Therefore,
if $G^{R}$ satifies the Usadel equation with boundary conditions, then also $YG^{R}Y$ does. Therefore, $G^{R} = YG^{R}Y$. This means that $G^{R}$ can be parametrised as:
\begin{equation}
    G^{R} = \begin{bmatrix}\cosh{\theta_{\uparrow}}&0&0&\sinh{\theta_{\uparrow}}e^{i\chi_{\uparrow}}\\0&\cosh{\theta_{\downarrow}}&\sinh{\theta_{\downarrow}}e^{i\chi_{\downarrow}}&0\\0&-\sinh{\theta_{\downarrow}e^{-i\chi_{\downarrow}}}&-\cosh{\theta_{\downarrow}}&0\\-\sinh{\theta_{\uparrow}}e^{-i\chi_{\uparrow}}&0&0&-\cosh{\theta_{\uparrow}}\end{bmatrix},
\end{equation}
In this parametrisation the normalisation condition $(G^{R})^{2} = \mathbf{1}$ is automatically satisfied. Moreover, the equations for $\theta_{\uparrow}, \chi_{\uparrow}$ are completely decoupled and very similar. They read
\begin{align}
    &\dv{^{2}\theta_{\uparrow,\downarrow}}{x^{2}}+2i\eta E\sinh{\theta_{\uparrow,\downarrow}} =0,\\
    &\dv{}{x}\left(\sinh^{2}{\theta_{\uparrow,\downarrow}}\dv{\chi_{\uparrow,\downarrow}}{x}\right) = 0.\\
\end{align}
The second of these equations represents supercurrent conservation \cite{chandrasekhar2003introduction}. In an SNN junction no supercurrent can flow \cite{belzig1999quasiclassical}, therefore the second equation reduces to $\chi_{\uparrow,\downarrow}$ being constant, the constant to be determined by the boundary condition at $x = 0$.\\
The boundary conditions for $\theta_{\uparrow,\downarrow}$ at $x = L$ are given by
\begin{equation}
    \theta_{\uparrow,\downarrow}(L) = 0.
\end{equation}
The boundary condition at $x = 0$ will be discussed in the coming sections.
In what comes, the spin notation will be suppressed. Focus will be on the spin up component, the discussion of the spin down component is completely analogous, with the spin singlet pair potential $\Delta_{s}$ reversing sign, while the triplet pair potential $\Delta_{p}$ is identical for both spin components.\\
For the Keldysh component, a parametrisation is used \cite{belzig1999quasiclassical}, \cite{chandrasekhar2003introduction}, which ensures that the normalisation condition is automatically satisfied:
\begin{align}
    G^{K}&=G^{R}h-hG^{A},\\
    h& = \begin{bmatrix}f_{L\uparrow}+f_{T\uparrow}&0&0&0\\0&f_{L\downarrow}+f_{T\downarrow}&0&0\\0&0&f_{L\downarrow}-f_{T\downarrow}&0\\0&0&0&f_{L\uparrow}-f_{T\uparrow}\end{bmatrix},\\
    f_{L\uparrow,\downarrow},f_{T\uparrow,\downarrow}&\in\mathbb{R}.
\end{align}
In an SNN junction, the Keldysh equation can be solved analytically if the solution to the retarded part is known, because the spectral supercurrent vanishes. A detailed discussion of the Keldysh equation and the calculation of the conductance using the Keldysh equation can be found in supplementary material \ref{sec:NEWconductanceformula}.

\section{Decoupling of equations}\label{sec:Decoupling}
In this section it is shown that for the s + helical p-wave junction the Usadel equation for the retarded part can be decomposed into two equations for two by two matrices, even though the direction of the d-vector in the superconductor depends on the direction of momentum.
First, the Green's function in the superconductor is discussed, and it is shown that inside the superconductor the spin sectors can not be separated. Afterwards it is shown that still though, in the normal metal the spin sectors can be separated. This has the  advantage that the $\theta$-parametrisation for two by two matrices can be used for both spin-sectors, which is much faster than solving the full four by four matrix equation.\\
As discussed in the previous sections, if the $d$-vector points in the z-direction, and $\Delta_{s,p}$ are real, the Green's function in the superconductor is given by
\begin{align}
    G^{R} &= \begin{bmatrix}
    \frac{E}{A_{1}}&0&0&\frac{\Delta_{+}}{A_{1}}\\0&\frac{E}{A_{2}}&\frac{\Delta_{-}}{A_{2}}&0\\0&-\frac{\Delta_{-}}{A_{2}}&\frac{E}{A_{2}}&0\\-\frac{\Delta_{+}}{A_{1}}&0&0&\frac{E}{A_{1}}
    \end{bmatrix}\\
    A_{1}& = \sqrt{E^{2}-\Delta_{+}^{2}},
    A_{2} = \sqrt{E^{2}-\Delta_{-}^{2}},\\
    \Delta_{+} &= \Delta_{s}+\Delta_{p} \label{eq:Deltaplus}\\
    \Delta_{-} &= \Delta_{s}-\Delta_{p} \label{eq:Deltamin}.
\end{align}
The Green's function in case the d-vector points in another direction, such as $(\cos{\phi},\sin{\phi},0)$ can be found by a rotation in spin space. To obtain the Green's function for $d = (\cos{\phi},\sin{\phi},0)$. The spin rotation matrix is $W_{1}(\phi) = \frac{1}{\sqrt{2}}(\cos{\phi}\sigma_{x}+\sin{\phi}\sigma_{y}+\sigma_{z})$. Applying this spin rotation matrix by $\Tilde{G}_{S}^{R}(\phi) = W(\phi)G^{R}W(\phi)$, where $W(\phi) = \begin{bmatrix}W_{1}(\phi)&0\\0&\sigma_{y}W_{1}(\phi)\sigma_{y}
\end{bmatrix}$, it is found that the Green's function in an s + helical p-wave superconductor is given by
\begin{align}\label{eq:GSphi1}
    \Tilde{G}_{S}^{R}(\phi) &= \begin{bmatrix}E(\frac{1}{A_{1}}+\frac{1}{A_{2}})&Ee^{-i\phi}(\frac{1}{A_{1}}-\frac{1}{A_{2}})&-e^{-i\phi}(\frac{\Delta_{+}}{A_{1}}-\frac{\Delta_{-}}{A_{2}})&\frac{\Delta_{+}}{A_{1}}+\frac{\Delta_{-}}{A_{2}}\\Ee^{i\phi}(\frac{1}{A_{1}}-\frac{1}{A_{2}})&E(\frac{1}{A_{1}}+\frac{1}{A_{2}})&-(\frac{\Delta_{+}}{A_{1}}+\frac{\Delta_{-}}{A_{2}})&e^{i\phi}(\frac{\Delta_{+}}{A_{1}}-\frac{\Delta_{-}}{A_{2}})\\e^{i\phi}(\frac{\Delta_{+}}{A_{1}}-\frac{\Delta_{-}}{A_{2}})&\frac{\Delta_{+}}{A_{1}}+\frac{\Delta_{-}}{A_{2}}&-E(\frac{1}{A_{1}}+\frac{1}{A_{2}})&Ee^{i\phi}(\frac{1}{A_{1}}-\frac{1}{A_{2}})\\-(\frac{\Delta_{+}}{A_{1}}+\frac{\Delta_{-}}{A_{2}})&-e^{-i\phi}(\frac{\Delta_{+}}{A_{1}}-\frac{\Delta_{-}}{A_{2}})&Ee^{-i\phi}(\frac{1}{A_{1}}-\frac{1}{A_{2}})&-E(\frac{1}{A_{1}}+\frac{1}{A_{2}})\end{bmatrix},
\end{align}
Thus, for the Green's function in the superconductor,  $\sigma_{x}$, $\sigma_{y}$ and $\sigma_{z}$-components are all present, and with different coefficients for different $\phi$. This shows that in the superconductor there is no separations in spin sectors possible.\\
However, the normal metal is only influenced by the superconductor via the appearance of $\langle S^{R}(\phi)\rangle$ in the boundary term defined in section \ref{sec:SNNBC}, which is an angular averaged quantity. The impossibility to separate the spin sectors in the superconductor therefore does not imply that it is impossible in the normal metal as well.
Moreover, in the Usadel equation there are no spin-dependent terms, so if $\langle S^{R}(\phi)\rangle$ can be separated into two blocks of two by two matrices, so can the Green's function in the normal metal.\\
In the remainder of this section it is shown that if a Green's function $G$ satisfies $YGY = G$, where $Y$ is the matrix defined in section \ref{sec:Superconductivityandspin}, then the output $I^{R} = \langle S^{R}(\phi)\rangle$ of the Tanaka-Nazarov boundary condition satisfies $I_{R} = YI_{R}Y$. To achieve this, it will be shown that in fact $S^{R}(-\phi) = YS^{R}(\phi)Y$.\\
As it turns out, it is convenient to perform a basis transformation on the Green's function in \ref{eq:GSphi1}. The reason for this it that the matrix $Y$ implies independence of the Green's function for spins pointing in the z-direction. Here, the d-vector is given by $d = (\cos{\phi},\sin{\phi},0)$. Angular averaging over $(-\frac{\pi}{2},\frac{\pi}{2})$ leaves only the x-direction, indicating that a transformation to a basis of x-spins is needed.
Defining $G_{S}^{R}(\phi) = W_{0}\Tilde{G}_{S}^{R}(\phi)W_{0}$, where $W_{0} = \frac{1}{\sqrt{2}}\begin{bmatrix}\sigma_{x}+\sigma_{z}&0\\0&\sigma_{y}(\sigma_{x}+\sigma_{z})\sigma_{y}\end{bmatrix}$ it follows that:
\begin{align}
    G_{S}^{R}(\phi) &= \frac{1}{2}D_{1}+\frac{1}{2}\cos{\phi}D_{2}+\frac{i}{2}\sin{\phi}D_{3},\\
    D_{1}&=\begin{bmatrix}(\frac{E}{A_{1}}+\frac{E}{A_{2}})\mathbf{1}&-(\frac{\Delta_{+}}{A_{1}}+\frac{\Delta_{-}}{A_{2}})i\sigma_{y}\\-(\frac{\Delta_{+}}{A_{1}}+\frac{\Delta_{-}}{A_{2}})i\sigma_{y}&-(\frac{E}{A_{1}}+\frac{E}{A_{2}})\mathbf{1}\end{bmatrix}\\
    D_{2}&=\begin{bmatrix}(\frac{E}{A_{1}}-\frac{E}{A_{2}})\sigma_{z}&-(\frac{\Delta_{+}}{A_{1}}-\frac{\Delta_{-}}{A_{2}})\sigma_{x}\\(\frac{\Delta_{+}}{A_{1}}-\frac{\Delta_{-}}{A_{2}})\sigma_{x}&-(\frac{E}{A_{1}}-\frac{E}{A_{2}})\sigma_{z}\end{bmatrix}\\
    D_{3}&=\begin{bmatrix}i(\frac{E}{A_{1}}-\frac{E}{A_{2}})\sigma_{y}&-(\frac{\Delta_{+}}{A_{1}}-\frac{\Delta_{-}}{A_{2}})\mathbf{1}\\(\frac{\Delta_{+}}{A_{1}}-\frac{\Delta_{-}}{A_{2}})\mathbf{1}&-i(\frac{E}{A_{1}}-\frac{E}{A_{2}})\sigma_{y}\end{bmatrix}\\
    A_{1} &= \sqrt{E^{2}-\Delta_{+}^2}, A_{2} = \sqrt{E^{2}-\Delta_{-}^{2}}.
\end{align}
The three matrices $D_{1}$, $D_{2}$, $D_{3}$ satisfy the relations
\begin{align}
    D_{1}^{2} = d_{1}\mathbf{1}, D_{2}^{2}& = d_{2}\mathbf{1},
    D_{3}^{2} = d_{3}\mathbf{1}\\
    d_{1}  =2(1+\frac{E^{2}-\Delta_{+}\Delta_{-}}{A_{1}A_{2}}),d_{2}  &= 2(1-\frac{E^{2}-\Delta_{+}\Delta_{-}}{A_{1}A_{2}}) = -d_{3}\\ 
    D_{1}D_{3}+D_{3}D_{1}& = 0 = D_{2}D_{3}+D_{3}D_{2}\\
    YD_{1}Y = D_{1},
    YD_{2}Y &= D_{2},
    YD_{3}Y = -D_{3}
\end{align}
From this the quantities $H_{\pm} = \frac{1}{2}(G_{S}^{R}(\phi)\pm G_{S}^{R}(\pi-\phi))$ and $C^{R} = (H_{+}^{R})^{-1}-(H_{+}^{R})^{-1}H_{-}^{R}$ can be calculated: 
\begin{align}
    H_{+}^{R} &= \frac{1}{2}D_{1}+\frac{i}{2}\sin\phi D_{3}\\
    (H_{+}^{R})^{-1} &= \frac{4}{d_{1}-d_{3}\sin^{2}\phi}H_{+}\\
    H_{-}^{R} & = \frac{1}{2}\cos{\phi}D_{2}\\
    C^{R}(\phi)&=\frac{2}{d_{1}-d_{3}\sin^{2}\phi}(D_{1}-\frac{1}{2}\cos{\phi}D_{1}D_{2}+i\sin{\phi}D_{3}-\frac{1}{2}i\cos{\phi}\sin{\phi}D_{3}D_{2}).
\end{align}
Now, using the matrix $Y$ from equation \ref{eq:Ydefinition}, it follows that $YD_{3}D_{2}Y = D_{3}YD_{2}Y = -D_{3}D_{2}$.
From this it can be concluded that 
\begin{equation}
    C^{R}(-\phi) = YC^{R}(\phi)Y,
\end{equation}
It now follows from substitution in the Tanaka-Nazarov boundary conditions that
\begin{equation}
    S^{R}(-\phi) = YS^{R}(\phi)Y. 
\end{equation}
This means that the total current satisfies
\begin{align}
    \langle S^{R}\rangle &= \int_{-\frac{\pi}{2}}^{\frac{\pi}{2}}S^{R}(\phi)d\phi = \int_{0}^{\frac{\pi}{2}}\left(S^{R}(\phi)+S^{R}(-\phi)\right)d\phi=\int_{0}^{\frac{\pi}{2}}\left( S^{R}(\phi)+YS^{R}(\phi)Y\right)d\phi,
\end{align}
and hence $Y\langle S^{R}\rangle Y = \langle S^{R}\rangle$. This concludes the proof, the Green's function in the normal metal can be separated into two two by two blocks, similar to the chiral p-wave superconductor. The result is in fact even stronger. Taking into account the basis transformation, this separation is a separation of spins pointing in the x-direction in the original basis, that is, in the direction of the direction of the d-vector averaged over the half-angle $(-\frac{\pi}{2},\frac{\pi}{2})$, what will be called the 'halfangular averaged d-vector'. This statement is general, the separation of the spins occurs for spins pointing in the halfangular averaged d-vector. \\
For the implementation of the boundary condition, it is convenient to build the four by four matrix $G^{R}$ in the normal metal from $\theta_{\uparrow,\downarrow},\chi_{\uparrow,\downarrow}$ and then compute $\Tilde{S}$ as a four by four matrix, from which the two by two matrices for the different spin sectors can be extracted from the nonzero elements of $\langle S\rangle$ as $\begin{bmatrix}
S_{11}&S_{14}\\S_{41}&S_{44}
\end{bmatrix}$ and $\begin{bmatrix}
S_{22}&S_{23}\\S_{32}&S_{33}
\end{bmatrix}$, where the notation $S_{ij}$ is used to denote the $(i,j)$-component of $\langle S\rangle$.

\subsection{Relation between the different components}\label{sec:Qsymm}
It should be noted that for the calculation of $S(\phi)$ still four by four matrices are required. This means that the two systems of equations can in general not be solved separately, they are coupled. However, in this section it is shown that further simplification is possible. First, note that
\begin{align}
    D_{1}D_{2} = \frac{2E(\Delta_{-}-\Delta_{+})}{A_{1}A_{2}}
    \begin{bmatrix}0&\sigma_{x}\\\sigma_{x}&0\end{bmatrix},
    D_{3}D_{2} = \frac{2(E^{2}-\Delta_{+}\Delta_{-})}{A_{1}A_{2}}\begin{bmatrix}\sigma_{x}&0\\0&-\sigma_{x}\end{bmatrix}.
\end{align}
Now denote
\begin{align}
    Q &=\begin{bmatrix}0&\mathbf{1}\\-\mathbf{1}&0\end{bmatrix},\\
    G_{Q} &= QGQ.
\end{align}
The matrices $D_{1}$, $D_{3}$, $D_{1}D_{2}$ and $D_{2}D_{3}$ all satisfy $QMQ = M$. Therefore, it holds that
\begin{equation}
    C = QCQ.
\end{equation}
Moreover, $Q$ anti commutes with $\tau_{3}$, and since it holds that $Q^{2} = -\mathbf{1}$, and $Q$ is independent of position, it follows that $Q(G\nabla G)Q = -QGQQ\nabla GQ = -G_{Q}\nabla G_{Q}$. Thus,
\begin{equation}
    G_{Q}\nabla(G_{Q}) = -G\nabla G = iE[\tau_{3},G] = -iE[\tau_{3},G_{Q}],
\end{equation}
that is, if $Q$ satisfies the Usadel equation in the bar, then $G_{Q}$ does so.
Thus, $G_{Q}$ satisfies the same equation combined with the same boundary condition, since $C = QCQ$. But this implies that $G_{Q}$, that is, also $G$ satisfies $QGQ = G$. This means that
\begin{align}
    \cosh{\theta_{\downarrow}} &= \cosh{\theta_{\uparrow}},
    \sinh{\theta_{\downarrow}}e^{i\chi\downarrow} =-\sinh{\theta_{\uparrow}}e^{-i\chi_{\uparrow}},
    -\sinh{\theta_{\downarrow}}e^{-i\chi\downarrow} =\sinh{\theta_{\uparrow}}e^{i\chi_{\uparrow}},
\end{align}
which is solved by $\theta_{\downarrow} = -\theta_{\uparrow}$, $\chi_{\downarrow} = -\chi_{\uparrow}$, similar to the cases studied before. Thus, only one system of two equations needs to be solved, only the equation for $\theta_{\uparrow},\chi_{\uparrow}$. Thus, from the computed $\langle S\rangle$, only $\begin{bmatrix}
S_{11}&S_{14}\\S_{41}&S_{44}
\end{bmatrix}$ needs to be extracted. For an SNN junction this means that only a differential equation for $\theta$ and an algebraic equation for $\chi$ need to be solved \cite{tanaka2021phys}. 
\subsection{Low energy behaviour for s + helical p-wave superconductors}
To investigate the low energy behaviour of the junction with s + helical p-wave superconductors, note that for each $\phi$ a basis transformation can be performed on the spin so that the spin points in the z-direction in the new frame, that is, in this basis the matrix $\cos{\phi}\sigma_{x}+\sin{\phi}\sigma_{y}$ becomes $\sigma_{z}$. 
In this basis the Green's function is given by
\begin{equation}
    \Bar{G} = \begin{bmatrix}
    \frac{E}{\sqrt{E^{2}-|\Delta_{+}|^{2}}}&0&0&\frac{\Delta_{+}}{\sqrt{E^{2}-|\Delta_{+}|^{2}}}\\
    0&\frac{E}{\sqrt{E^{2}-|\Delta_{-}|^{2}}}&-\frac{\Delta_{-}}{\sqrt{E^{2}-|\Delta_{-}|^{2}}}&0\\
    0&\frac{\Delta_{-}}{\sqrt{E^{2}-|\Delta_{-}|^{2}}}&\frac{E}{\sqrt{E^{2}-|\Delta_{-}|^{2}}}&0\\
    -\frac{\Delta_{+}}{\sqrt{E^{2}-|\Delta_{+}|^{2}}}&0&0&\frac{E}{\sqrt{E^{2}-|\Delta_{+}|^{2}}}
    \end{bmatrix},
\end{equation}
where
\begin{align}
    \Delta_{\pm} = \Delta\frac{1\pm r}{\sqrt{r^{2}+1}}.
\end{align}
Thus in this basis, the low energy limit is given by
\begin{align}
    C(E<<\Delta)\approx
    \begin{cases}
    \begin{bmatrix}
    0&0&0&-i\\
    0&0&i&0\\
    0&-i&0&0\\
    i&0&0&0
    \end{bmatrix}&r<1\\\frac{i}{E}\frac{1-r^{2}}{\sqrt{r^{2}+1}}\begin{bmatrix}
    1&0&0&i\\
    0&1&i&0\\
    0&i&-1&0\\
    i&0&0&-1
    \end{bmatrix}&r>1.
    \end{cases}
\end{align}
 For $r<1$ the expression for $C$ converges as the energy goes to zero. On the other hand, the expression for $C$ at low energies for $r>1$ has a prefactor that diverges at low energies. Moreover, seemingly contradictory, it does not satisfy the normalisation condition, since it squares to the zero matrix. The reason for this is that only the leading order in energy (here -1) is shown here, it is the same approximation as $|\cosh{\theta}| = |\sinh{\theta}|+O(|\theta|^{-1})$ for large $|\theta|$. Taking into account the higher order terms shows $C$ does satisfy the normalisation conditions. However, those higher order terms do not have any effect on the matrix current $I^{R}$. That the lower order terms do not give any extra contribution at low energies can be seen from the identity
\begin{align}
    \lim_{E\xrightarrow{}0}I^{R}&= \lim_{E\xrightarrow{}0}T_{1}(1+T_{1}^{2}+2T_{1}(CG+GC))^{-1}(CG-GC) =\nonumber\\& \lim_{E\xrightarrow{}0}T_{1}(E+T_{1}^{2}E+2T_{1}(ECG+GEC))^{-1}(ECG-GEC) = (\Tilde{C}G+G\Tilde{C})^{-1}(\Tilde{C}G-G\Tilde{C}),
\end{align}
where
\begin{align}
    \Tilde{C} = \lim_{E\xrightarrow{}0}EC = \begin{bmatrix}
    1&0&0&i\\
    0&1&i&0\\
    0&i&-1&0\\
    i&0&0&-1
    \end{bmatrix}
\end{align}
is independent of the actual value of $r$.
Thus, in this basis, the zero energy limit only takes two values, depending on whether $r<1$ or $r>1$. Since equality is basis independent, it follows that at zero energy the solution is constant in the ranges $(0,1)$ and $(1,\infty)$. 
\subsection{Explicit expressions}
To obtain explicit expressions for the components $S_{ij}$ of $\langle S\rangle$ where $i,j\in\{1,2,3,4\}$, the following procedure was followed. In the following, to shorten notation, $\Omega_{\pm} = \sqrt{E^{2}-\Delta_{\pm}^{2}}$ is used.
First, $\langle S\rangle$ is expressed as
\begin{align}
    &\langle S \rangle = \int_{0}^{\frac{\pi}{2}}\left(S(\phi)+S(-\phi)\right)d\phi\\
    &S(\phi) = D^{-1}N\\&D =((4-2T)+T(CG+GC))\\&N = (CG-GC).
\end{align}
Note that $C = C_{1}+C_{2}$, where
\begin{equation}
    C_{1} = \frac{2}{d_{1}-d_{3}\sin{\phi}}D_{1}(1-\frac{1}{2}\cos{\phi}D_{2}),
    C_{2} = \frac{2}{d_{1}-d_{3}\sin{\phi}}i\sin{\phi}D_{3}(1-\frac{1}{2}\cos{\phi}D_{2}),
\end{equation}
where $C_{1,2}$ satisfy $YC_{1}Y = C_{1}$ and $YC_{2}Y = C_{2}$.
It can be explicitly calculated by plugging in $G = \begin{bmatrix}
\cosh{\theta}&0&0&\sinh{\theta}e^{i\chi}\\0&\cosh{\theta}&-\sinh{\theta}e^{-i\chi}&0\\0&\sinh{\theta}e^{i\chi}&-\cosh{\theta}&0\\-\sinh{\theta}e^{-i\chi}&0&0&-\cosh{\theta}
\end{bmatrix}$ that
\begin{align}
    C_{1}G+GC_{1} &= \alpha\mathbf{1}\\
    \alpha &=\frac{1}{2}\frac{\Omega_{+}\Omega_{-}}{\sqrt{(E^{2}-\Delta_{+}\Delta_{-})(2(E^{2}-\Delta_{+}\Delta_{-}-\Omega_{+}\Omega_{-}))}} \Tilde{\alpha}\\
    \Tilde{\alpha} &= 2E\Big(\frac{1}{\Omega_{+}}+\frac{1}{\Omega_{-}}\Big)\cosh{\theta}+\Big(\frac{\Delta_{+}}{\Omega_{+}}+\frac{\Delta_{-}}{\Omega_{-}}\Big)\sin{\theta}\cos{\chi}\nonumber\\&+2iE\Big(\frac{\Delta_{+}-\Delta_{-}}{\Omega_{+}\Omega_{-}}\Big)\sinh{\theta}\sin{\chi}
\end{align}
Denoting $\Tilde{C}_{2} = C_{2}G+GC_{2}$, it follows that
\begin{align}
    D(\phi)^{-1}+D(-\phi)^{-1} = (4-2T+\alpha T+\Tilde{C_{2}})^{-1}+(4-2T+\alpha T-\Tilde{C_{2}})^{-1} = 2\alpha((4-2T+\alpha T)^{2}-\Tilde{C}_{2}^{2})^{-1}
\end{align}
and
\begin{align}
    D(\phi)^{-1}-D(-\phi)^{-1} = (4-2T+\alpha T+\Tilde{C}_{2})^{-1}-(4-2T+\alpha T-\Tilde{C_{2}})^{-1} = 2((4-2T+\alpha T)^{2}-\Tilde{C}_{2}^{2})^{-1}\Tilde{C}_{2}
\end{align}
Now, $\Tilde{C_{2}}^{2} = \beta\mathbf{1}$ with
\begin{align}
\beta&= \frac{1}{4}\frac{\Omega_{+}\Omega_{-}}{E^{2}-\Delta_{+}\Delta_{-}}\Tilde{\beta}\\
    \Tilde{\beta} &= |\frac{-2(E^{2}-\Delta_{+}\Delta_{-})+2\Omega_{+}\Omega_{-}-E\big(\Omega_{+}-\Omega_{-}\big)}{\Omega_{+}^{2}\Omega_{-}^{2}}\cosh{\theta}\nonumber\\&+\frac{\Delta_{-}\Omega_{+}-\Delta_{+}\Omega_{-}}{\Omega_{+}^{2}\Omega_{-}^{2}}\sinh{\theta}e^{i\chi}|^{2}-\sinh^{2}{\theta}\Bigg(2(\Delta_{+}\Delta_{-}-E^{2}+\Omega_{+}\Omega_{-})\cos{\chi}\nonumber\\&+iE(\Omega_{+}+\Omega_{-})\sin{\chi}\Bigg)^{2}
\end{align}

Using this, it follows that
\begin{align}
    S(\phi)+S(-\phi) &= D(\phi)^{-1}[C_{1}+C_{2},G]+(D(\phi))^{-1}[C_{1}-C_{2},G]\\
    &=2\alpha((4-2T+\alpha T)^{2}-\Tilde{C}_{2}^{2})^{-1}[C_{1},G]+2((4-2T+\alpha T)^{2}-\Tilde{C}_{2}^{2})^{-1}\Tilde{C}_{2}[C_{2},G]
\end{align}

Denoting $\Tilde{S}_{a} = [C_{1},G_{1}]$ and $\Tilde{S}_{b} = \Tilde{C}_{2}[C_{2},G]$ it follows that
\begin{align}\Tilde{S}_{a} &= \begin{bmatrix}S_{a11}&0&0&S_{a14}\\0&S_{a22}&S_{a23}&0\\0&S_{a32}&S_{a33}&0\\S_{a41}&0&0&S_{a44}\end{bmatrix},\\
S_{a11} &= 2\sinh{\theta}\frac{2E(\Delta_{+}-\Delta_{-})\cos{\chi}+i(\Delta_{+}\Omega_{-}+\Delta_{-}\Omega_{+})\sin{\chi}}{\Omega_{+}\Omega_{-}}\\
S_{a14}&=2\frac{\Big(\Delta_{-}(2E+\Omega_{+})-\Delta_{+}(2E-\Omega_{-})\Big)\cosh{\theta}+E\big(\Omega_{+}+\Omega_{-}\big)\sinh{\theta}e^{i\chi}}{\Omega_{+}\Omega_{-}}\\
S_{a41}&=2\frac{\Big(\Delta_{-}(-2E+\Omega_{+})+\Delta_{+}(-2E+\Omega_{-})\Big)\cosh{\theta}+E\big(\Omega_{+}+\Omega_{-}\big)\sinh{\theta}e^{i\chi}}{\Omega_{+}\Omega_{-}}\\
S_{a44}& = -S_{a11}\\
S_{a22} &= -S_{a11}\\
S_{a23}&=-S_{a41}\\
S_{a32}&=-S_{a14}\\
S_{a33}& = -S_{a22}
\end{align} 
whereas $S_{b}$ is given by
\begin{align}\Tilde{S}_{b}& = \begin{bmatrix}S_{b11}&0&0&S_{b14}\\0&S_{b22}&S_{b23}&0\\0&S_{b32}&S_{b33}&0\\S_{b41}&0&0&S_{b44}\end{bmatrix},\\
S_{b11} & = \frac{4\sinh{\theta}}{\Omega_{+}^{2}\Omega_{-}^{2}}\Bigg(2(E^{2}-\Delta_{+}\Delta_{-}-\Omega_{+}\Omega_{-})\cos{\chi}+iE(\Omega_{-}-\Omega_{+})\sin{\chi}\Bigg)\nonumber\\&\Bigg((\Delta_{+}\Omega_{-}-\Delta_{-}\Omega_{+})\cosh{\theta}+E(\Omega_{-}-\Omega_{+})\sinh{\theta}\cos{\chi}\nonumber\\&+2i(E^{2}-\Delta_{+}\Delta_{-}-\Omega_{+}\Omega_{-})\sinh{\theta}\sin{\chi}\Bigg)\\
S_{b14} & = \frac{4}{(E^{2}-\Delta_{+}^{2})(E^{2}-\Delta_{-}^{2})}\Bigg(\Big(2(E^{2}-\Delta_{+}\Delta_{-})+E\Omega_{+}-E\Omega_{-}-2\Omega_{-}\Omega_{+}\Big)\cosh{\theta}\nonumber\\&+\left(\Delta_{+}\Omega_{-}-\Delta_{-}\Omega_{+}\right)\sinh{\theta}e^{i\chi}\Bigg)\nonumber\\&\Bigg(\left(\Delta_{+}\Omega_{-}-\Delta_{-}\Omega_{+}\right)\cosh{\theta}+E\left(\Omega_{-}-\Omega_{+}\right)\sinh{\theta}\cos{\chi}\nonumber\\&+2i\left(E^{2}-\Delta_{+}\Delta_{-}-\Omega_{+}\Omega_{-}\right)\sinh{\theta}\sin{\chi}\Bigg)\\
S_{b41}&=\frac{4}{\Omega_{+}^{2}\Omega_{-}^{2}}\Bigg(\Big(2\Omega_{+}^{2}\Delta_{-})-E\Omega_{+}+E\Omega_{-}-2\Omega_{+}\Omega_{-}\Big)\cosh{\theta}\nonumber\\&+\left(-\Delta_{+}\Omega_{-}-\Delta_{-}\Omega_{+}\right)\sinh{\theta}e^{-i\chi}\Bigg)\nonumber\\&\Bigg(\left(\Delta_{+}\Omega_{-}-\Delta_{-}\Omega_{+}\right)\cosh{\theta}+E\left(\Omega_{-}-\Omega_{+}\right)\sinh{\theta}\cos{\chi}\nonumber\\&+2i\left(E^{2}-\Delta_{+}\Delta_{-}-\Omega_{+}\Omega_{-}\right)\sinh{\theta}\sin{\chi}\Bigg)\\
S_{b44}&=-S_{b11}\\
S_{b22}& = -S_{b11}\\
S_{b23} & = -S_{b41}\\
S_{b32} &= -S_{b14}\\
S_{b33}&=-S_{b22}.
\end{align}
The elements of $S$ are now given by
\begin{equation}
    S_{ij} = \frac{2}{(4-2T+\alpha T)^{2}-\beta}(\alpha S_{aij}+S_{bij})
\end{equation}

\section{Solution procedure}
\subsection{The phase equation in two dimensions}\label{sec:chiNUM2D}
The expression for the potentially complex phase $\chi$ in the junction is a more difficult problem for quasi-one-dimensional junctions than for one-dimensional junctions. For one-dimensional junctions an analytic expression was found by requiring the $\tau_{3}$-component of $I$ to be 0 \cite{tanaka2021phys}. In quasi-one-dimensional junctions this is also possible, however, the current is now an average over the nodes. The resulting expression is equation \ref{eq:chiequation2D}:
\begin{align}
 &\int_{-\frac{\pi}{2}}^{\frac{\pi}{2}}\Big((1+T_{1}^{2})(1+g_{+}g_{-}  -f_{+}f_{-})+T_{1} ((g_{+}+g_{-})\cosh{\theta}\nonumber\\&- (f_{+}+f_{-}-(g_{-}f_{+}-g_{+}f_{-}) )e^{i(\psi-\chi)}\sinh{\theta}- (f_{+}+f_{-}+(g_{-}f_{+}-g_{+}f_{-}) )e^{-i(\psi-\chi)}\sinh{\theta} )\Big)^{-1}\nonumber\\&\left(e^{i(\psi-\chi)}(f_{+}+f_{-}-(g_{-}f_{+}-g_{+}f_{-})) - e^{i(\chi-\psi)}(f_{+}+f_{-}+(g_{-}f_{+}-g_{+}f_{-}))\right)T\cos{\phi} d\phi \nonumber \\&=0,\label{eq:chiequation2D}
\end{align}
where
\begin{align}
    g_{\pm}(\phi) &= \frac{E}{\sqrt{E^{2}-\Delta_{\pm}(\phi)}}\\
    f_{\pm}(\phi) &= \frac{\Delta_{\pm}(\phi)}{\sqrt{E^{2}-\Delta_{\pm}(\phi)}}\\
    \Delta_{\pm}(\phi)& \frac{1}{\sqrt{r^{2}+1}}\pm\frac{r}{\sqrt{r^{2}+1}}\cos{\phi},\\
    T(\phi) &= \frac{\cos^{2}\phi}{\cos^{2}{\phi}+(\frac{z}{2})^{2}},\\
    T_{1}(\phi) &= \frac{T(\phi)}{2-T(\phi)+2\sqrt{1-T(\phi)}}.
\end{align}
A big difference with the expression for the one-dimensional junction is that the phase in the junction depends on the parameter $\theta$. Instead of finding the analytical solution to equation \ref{eq:chiequation2D}, a numerical procedure was developed. Denote the integral in equation \ref{eq:chiequation2D} by $S_{\phi}$. The condition can then be shortly written as 
\begin{equation}
    S_{\phi}(\theta,\chi) = 0.
\end{equation}
Now start from an initial guess $\chi_{i}$ and denote the new guess $\chi_{i+1} = \chi_{i}+\delta\chi$. Then $\delta\chi$ is chosen such that
\begin{equation}
    S_{\phi}(\theta,\chi_{i+1})\approx S_{\phi}(\theta,\chi_{i})+\dv{S_{\phi}}{\chi}\delta\chi = 0,
\end{equation}
that is
\begin{equation}
    \delta\chi = -\frac{1}{S_{\phi}(\theta,\chi)}\dv{S_{\phi}}{\chi}.
\end{equation}
The quantity $\dv{S_{\phi}}{\chi}$ can be calculated to be
\begin{align}
    \dv{S_{\phi}}{\chi} =& 2T\left((2-T)\mathbf{1}+T(CG+GC)\right)^{-1}\left(C\dv{G}{\chi}-\dv{G}{\chi}C\right)\nonumber\\+&2T\left((2-T)\mathbf{1}+T(CG+GC)\right)^{-1}\left((2-T)\mathbf{1}+T(C\dv{G}{\chi}+\dv{G}{\chi}C)\right)\nonumber\\&\left((2-T)\mathbf{1}+T(CG+GC)\right)^{-1}\left(CG-GC\right),\\
    \dv{G}{\chi} =& \begin{bmatrix}0&i\sinh{\theta}e^{i\chi}\\i\sinh{\theta}e^{-i\chi}&0\end{bmatrix}.
\end{align}
This procedure was repeated until $S_{\phi}$ was lower than the tolerance. It was found that this procedure converges well if one takes for the first initial guess $\chi_{0}$ the phase in the one-dimensional junction using the same parameters, which can be calculated analytically \cite{tanaka2021phys}. Once there was solved for the parameter $\chi$, the parameter $\theta$ was recalculated by solving the Usadel equation using the same method as for one-dimensional junction.\\
This procedure was found to work well for $r<0.95$ and $r>1.5$. When $r$ is close to 1, the code has trouble with convergence. These convergence issues can be traced back to the one-dimensional case, where $r = 1$ can not be solved using the $\theta$-parametrisation, and where near $r = 1$ the imaginary part of $\chi$ can become large. For spin up electrons this imaginary part is positive, for spin down electrons this imaginary part is negative. Therefore, for the solution of $S_{\phi}(\theta,\chi) = 0$ the parameter $X_{\uparrow,\downarrow}$ was introduced:
\begin{align}
    X_{\uparrow} &= \sinh{\theta_{\uparrow}}e^{-i\chi}\\
    X_{\downarrow}  &=\sinh{\theta_{\downarrow}}e^{i\chi}\\
    G_{\uparrow}& = \begin{bmatrix}\cosh{\theta_{\uparrow}}&\frac{\sinh^{2}{\theta_{\uparrow}}}{X_{\uparrow}}\\-X_{\uparrow}&-\cosh{\theta_{\uparrow}}\end{bmatrix}\\
    G_{\downarrow}& = \begin{bmatrix}\cosh{\theta_{\downarrow}}&X_{\downarrow}\\-\frac{\sinh^{2}{\theta_{\downarrow}}}{X_{\downarrow}}&-\cosh{\theta_{\downarrow}}\end{bmatrix}.
\end{align}
Contrary to the parameter $\chi$ the parameter $X$ does not become very large. Using this procedure also $0.95<r<1$ and $1<r<1.5$ could be calculated.
\subsection{The density of states equation}
Whereas the parameter $\chi$ is solved analytically, the parameter $\theta$ which determines the density of states via $\nu = \nu_{0}\text{Re}(\cosh{\theta})$, where $\nu_{0}$ is the density of states in the normal state, is obtained using a numerical scheme. In this section the numerical scheme will be discussed. The discussion here will be general for both one-dimensional and quasi-one-dimensional junctions, as the method is very similar for those two cases. Recall that the equation to be solved for an SNN junction is
\begin{equation}
    \dv{^{2}\theta}{x^{2}}+2i\eta E\sinh{\theta} = 0,\\
    \dv{\theta}{x} = \frac{1}{\gamma_{B}L\langle T\rangle}\langle S_{\theta}\rangle,\\
    \theta(L) = 0,
\end{equation}
where $\eta = \frac{\Delta}{2\pi T_{c}}$ according to the BCS-theory \cite{SuzukiUsa}, $\gamma_{B} = \frac{R_{b}L}{R_{N}}$ is the ratio of boundary resistance to junction resistivity in the normal state, $T(\phi) = \frac{\cos^{2}(\phi)}{\cos^{2}(\phi)+(\frac{z}{2})^{2}}$ is the transparency and the notation $\langle\rangle$ is used for averaging over the angle $\phi\in(-\frac{\pi}{2},\frac{\pi}{2})$ that will be needed in quasi-one-dimensional junctions. In one-dimensional junctions there is only one mode with $\phi = 0$ and the angular averaging is trivial. The term $S_{\theta}$ is the Tanaka-Nazarov boundary term \cite{nazarov1999novel}, \cite{tanaka2003circuit}, \cite{tanaka2004theory} as written in compact form by Tanaka \cite{tanaka2021phys}:
\begin{align}
    S_{\theta} &= e^{-i\chi}S_{1,2}+e^{i\chi}S_{2,1},\\
    S &= 2T\left((2-T)\mathbf{1}+T(CG+GC)\right)^{-1}\left(CG-GC\right),\\
    G_{R}& = \begin{bmatrix}\cosh{\theta}&\sinh{\theta}e^{i\chi}\\-\sinh{\theta}e^{-i\chi}&-\cosh{\theta}\end{bmatrix},\\
    C^{R} & = (H_{+})^{-1}(\mathbf{1}-H_{-}),\\
    H_{\pm} &= \frac{1}{2}(G_{S}(\phi)\pm G_{S}(\pi-\phi)).\\
\end{align}
Here $S_{i,j}$ denotes the $(i,j)$-component of $S$, $\mathbf{1}$ is used to denote the identity matrix, $G_{S}$ is the now 2 by 2 Green's function in the superconductor, and $\chi$ is the phase in the normal metal at the interface with the superconductor.
Note that the $(\dv{\chi}{x})^{2}$-term in the equation \cite{belzig1999quasiclassical} does not appear in the Usadel equation as the phase in an SNN junction is constant \cite{belzig1999quasiclassical}.\\
The Usadel equation is a non-linear equation and is thus to be solved using an iterative scheme. The scheme used here is as follows:
\begin{itemize}
    \item Suppose we have a guess $\theta^{i}$ for the solution of the Usadel equation.
    \item Write $\theta^{i+1} = \theta^{i}+\delta\theta$, where $\theta^{i+1}$ is the updated guess for the solution of the Usadel equation.
    \item Using a Taylor expansion write $\sinh{\theta^{i+1}} \approx \sinh{\theta^{i}}+\cosh{\theta^{i}}\delta\theta$ and $S(\theta^{i+1})\approx S(\theta^{i})+\dv{S}{\theta}\delta\theta$.
    \item The equation for $\delta\theta$ becomes:
\begin{align}
    &\dv{^{2}\delta\theta}{x^2}+2iE\cosh{\theta^{i}}\delta\theta = -\dv{^{2}\theta^{i}}{x^{2}}-2iE\sinh{\theta^{i}},\label{eq:Usadeldelta}\\ 
    &\dv{\delta\theta}{x}+\dv{S}{\theta}\delta\theta =  -\dv{\theta^{i}}{x}-S(\theta^{i}),\label{eq:UsadeldeltaBC1}\\
    &\delta\theta(L) = 0.\label{eq:UsadeldeltaBC2}
\end{align}
    \item Solve for $\delta\theta$ and obtain new guess $\theta^{i+1}$.
\end{itemize}
The quantity $\dv{S}{\theta}$ can be calculated to be
\begin{align}
    \dv{S}{\theta} =& 2T\left((2-T)\mathbf{1}+T(CG+GC)\right)^{-1}\left(C\dv{G}{\theta}-\dv{G}{\theta}C\right)\nonumber\\+&2T\left((2-T)\mathbf{1}+T(CG+GC)\right)^{-1}\left((2-T)\mathbf{1}+T(C\dv{G}{\theta}+\dv{G}{\theta}C)\right)\nonumber\\&\left((2-T)\mathbf{1}+T(CG+GC)\right)^{-1}\left(CG-GC\right),\\
    \dv{G}{\theta} =& \begin{bmatrix}\sinh{\theta}&\cosh{\theta}e^{i\chi}\\-\cosh{\theta}e^{-i\chi}&\sinh{\theta}\end{bmatrix}.
\end{align}
This procedure was repeated until $\text{max}|\theta^{i+1}-\theta^{i}|$ becomes smaller than the tolerance, which was predetermined to be $10^{-12}$. This leaves two questions: what to use for the initial guess $\theta^{0}$ and how to solve for $\delta\theta$.\\
The code starts by solving the equation for $E = 3\Delta_{0}$, for which the proximity effect is small and $\theta^{0} = 0$ is a good initial guess. Then, it solves the equation for energy $E-dE$ by using the solution found at energy $E$. This is repeated until E becomes negative. The solutions for negative energies are calculated by starting at $E = -3\Delta$ and then stepping up in energy.\\
To find $\delta\theta$, equations \ref{eq:Usadeldelta} to \ref{eq:UsadeldeltaBC2} need to be solved. The grid is discretised using $N_{x}$ grid points $x_{j} = j\Delta x, j = 0...N_{x}-1$, with $\delta\theta_{j} = \delta\theta(x_{j})$ and $\theta^{i}_{j} = \theta^{i}(x_{j})$. These equations are linear in $\delta\theta$, and therefore the equation can be solved using MATLAB. For the bulk equation, the standard second order discretisation $\dv{^{2}\delta\theta}{x^{2}}|_{x = j\Delta x}\approx\frac{\delta\theta_{j+1}+\delta\theta_{j-1}-2\delta\theta_{j}}{(\Delta x)^{2}}$, where $\Delta x$ is the grid spacing, was used. This results in the equation
\begin{align}
    &\frac{\delta\theta_{1}+\delta\theta_{-1}-2\delta\theta_{0}}{\Delta x^2}+2iE\cosh{\theta^{i}_{0}} = -\frac{\theta^{i}_{1}+\theta^{i}_{-1}-2\theta^{i}_{0}}{\Delta x^2}-2iE\sinh{\theta^{i}},\label{eq:Usadelnum}\\
    &j = 1...N_{x}-2\nonumber
\end{align}

For the boundary condition on the left hand side a ghost cell method was used. That is, by introducing the quantity $\theta_{-1}$, which is never computed, for $j = 0$ we have 
\begin{align}
    &\frac{\delta\theta_{1}+\delta\theta_{-1}-2\delta\theta_{0}}{\Delta x^2}+2iE\cosh{\theta^{i}_{0}} = -\frac{\theta^{i}_{1}+\theta^{i}_{-1}-2\theta^{i}_{0}}{\Delta x^2}-2iE\sinh{\theta^{i}},\\
    &\frac{\delta\theta_{1}-\delta\theta_{-1}}{2\Delta x}-\pdv{S}{\theta}(\theta^{i}_{0})\delta\theta_{0} =-\frac{\theta^{i}_{1}-\theta^{i}_{-1}}{2\Delta x} -S(\theta^{i}_{0})
\end{align}
which can be used to find the single boundary condition
\begin{align}
    &2\frac{\delta\theta_{1}-\delta\theta_{0}}{\Delta x^2}+\frac{2}{\Delta x}\pdv{S}{\theta}(\theta^{i}_{0})\delta\theta_{0}+2iE\cosh{\theta^{i}_{0}} = -2\frac{\theta^{i}_{1}+-\theta^{i}_{0}}{\Delta x^2}-\frac{2}{\Delta x}S(\theta^{i}_{0})-2iE\sinh{\theta^{i}}.\label{eq:UsNumBC1}\\
\end{align}
This method solves the boundary condition to second order. The boundary condition on the right hand side is a homogeneous Dirichlet boundary condition, and can thus implemented exact by setting \begin{equation}\delta\theta_{j = N_{x}-1} = 0.\label{eq:UsNumBC2}\end{equation}
Together, equation \ref{eq:Usadelnum}, \ref{eq:UsNumBC1}, \ref{eq:UsNumBC2} form a linear system of equations that can be solved using MATLAB. 
\section{New conductance formula}\label{sec:NEWconductanceformula}
In this section, a new conductance formula will be derived. The derivation will follow a path very similar to the derivation presented in \cite{tanaka2021phys}. However, the assumption that the distribution functions for up and down spin are equal, which  is not made here. Namely, this assumption is valid both in pure s-wave and pure p-wave superconductors and in equilibrium mixed type superconductors, but not necessary in non-equilibrium mixed type superconductors due to the polarisation in $\Delta = \Delta_{s}+\Delta_{p}e^{i\phi}\sigma_{z}$. This is done to allow for possible differences in the distribution function for up and down spin in the normal metal when using mixed s + p-wave superconductors. It is found that if the superconductor is not a pure s-wave or pure p-wave superconductor, the distribution functions are in general unequal. First the general pair potential in which the d-vector points in the z-direction will be discussed, afterwards the helical case will be considered.\\
As shown in \cite{kokkeler2021usadel}, the Keldysh equations in absence of a supercurrent are
\begin{align}
    D_{L\uparrow,
    \downarrow}\nabla f_{L\uparrow,
    \downarrow} + C_{L\uparrow,
    \downarrow}\nabla f_{T\uparrow,
    \downarrow} &= \frac{1}{\gamma_{B}L}\text{Trace}(I^{K}),\label{eq:DLeq}\\
    D_{T\uparrow,
    \downarrow}\nabla f_{T\uparrow,
    \downarrow} + C_{T\uparrow,
    \downarrow}\nabla f_{L\uparrow,
    \downarrow} &= \frac{1}{\gamma_{B}L}\text{Trace}(\tau_{3}I^{K\uparrow,
    \downarrow})\label{eq:DTeq},\\
    f_{L\uparrow,
    \downarrow}(L) &= \frac{1}{2}(\tanh{\frac{E+eV}{kT}}+\tanh{\frac{E-eV}{kT}})\\
    f_{T\uparrow,
    \downarrow}(L) &= \frac{1}{2}(\tanh{\frac{E+eV}{kT}}-\tanh{\frac{E-eV}{kT}}),
\end{align}
where $f_{L}$ and $f_{T}$ are the distribution functions, $I^{K}$ is the Keldysh component of the Tanaka-Nazarov boundary condition, $\tau_{3} = \begin{bmatrix}1&0\\0&-1\end{bmatrix}$, $E$ is the energy, $e$ electrical charge, $V$ the voltage applied to the normal reservoir, $k$ the Boltzmann constant, and $T$ the temperature. The coefficients are given by
\begin{align}
    D_{L\uparrow,
    \downarrow} &= 1+|\cosh{\theta_{\uparrow,
    \downarrow}}|^{2}-|\sinh{\theta_{\uparrow,
    \downarrow}}|^{2}\cosh{2\text{Im}(\chi_{\uparrow,
    \downarrow})}\\
    D_{T\uparrow,
    \downarrow} &= 1+|\cosh{\theta_{\uparrow,
    \downarrow}}|^{2}+|\sinh{\theta_{\uparrow,
    \downarrow}}|^{2}\cosh{2\text{Im}(\chi_{\uparrow,
    \downarrow})}\\
    C_{\uparrow,\downarrow}: = C_{L\uparrow,
    \downarrow}& = C_{T\uparrow,
    \downarrow} = -|\sinh{\theta_{\uparrow,
    \downarrow}} |^{2}\sinh{2\text{Im}(\chi_{\uparrow,
    \downarrow})},
\end{align}
where $\theta_{\uparrow,\downarrow}$ and $\chi_{\uparrow,\downarrow}$ are the solutions from the retarded equation.\\
Now, as shown in \cite{tanaka2021phys} and as verified by the code, the solutions $\theta_{\uparrow, \downarrow}$ and $\chi_{\uparrow,\downarrow}$ satisfy
\begin{align}
    |\cosh{\theta_{\uparrow}}| &= |\cosh{\theta_{\downarrow}}|\\
    |\sinh{\theta_{\uparrow}}| &= |\sinh{\theta_{\downarrow}}|\\
    \sinh{\text{Im}(\chi_{\uparrow})} &= -\sinh{\text{Im}(\chi_{\downarrow})}. 
\end{align}
Therefore the coefficients satisfy
\begin{align}
    D_{L}&:= D_{L\uparrow} = D_{L\downarrow}\\
    D_{T}&:= D_{T\uparrow} = D_{T\downarrow}\\
    \Tilde{C}&:= C_{\uparrow} = -C_{\downarrow}.
\end{align}
Using this, by subtracting equation \ref{eq:DLeq} for different spin, and by adding equation \ref{eq:DTeq} for different spin, one obtains the equations
\begin{align}
    &D_{L}\nabla f_{L} + C_{L}\nabla f_{T} = \frac{1}{\gamma_{B}L}\text{Tr}(I^{K\uparrow}-I^{K\downarrow})\label{eq:fL}\\
    &D_{T}\nabla f_{T} + C_{T}\nabla f_{L} = \frac{1}{\gamma_{B}L}\text{Tr}((I^{K\uparrow}+I^{K\downarrow})\tau_{3}) \label{eq:fT},\\
    &f_{L}(L) = 0\label{eq:fLBC}\\
    &f_{T}(L) = (\tanh{\frac{E+eV}{kT}}-\tanh{\frac{E-eV}{kT}}) =:f_{T0}.\label{eq:fTBC}
\end{align}
In this expression
\begin{align}
    f_{L} &= f_{L\uparrow}-f_{L\downarrow}\\
    f_{T} & = f_{T\uparrow}+f_{T\downarrow}.
\end{align}
Now, according to \cite{tanaka2021phys}, the Keldysh part of the Tanaka-Nazarov boundary condition can be written as
\begin{align}
    \text{Tr}(I^{K\uparrow,\downarrow}\tau_{3})&=I_{b1\uparrow,\downarrow}f_{L\uparrow,\downarrow}(0)-I_{bs\uparrow,\downarrow}f_{S\uparrow,\downarrow}(0)+I_{b0\uparrow,\downarrow}f_{T\uparrow,\downarrow}(0),\\
    I_{b1\uparrow,\downarrow} &=\langle\frac{T_{1}}{|d_{R}|^{2}}\Bigg((1+T_{1}^{2})\text{Tr}\left(\Big((C+C^{\dagger})G+G^{\dagger}(C+C^{\dagger})\Big)\tau_{3}\right)\nonumber\\&+2T_{1}\text{Tr}\left(\Big(1+C^{\dagger}C+G^{\dagger}(1+C^{\dagger}C)G\Big)\tau_{3}\right)\Bigg)\rangle,\label{eq:Ib1expressionupdown}\\
    I_{bs\uparrow,\downarrow} &=\langle\frac{T_{1}}{|d_{R}|^{2}}\Bigg((1+T_{1}^{2})\text{Tr}\left(\Big((G+G^{\dagger})C+C^{\dagger}(G+G^{\dagger})\Big)\tau_{3}\right)\nonumber\\&+2T_{1}\text{Tr}\left(\Big(1+G^{\dagger}G+C^{\dagger}(1+G^{\dagger}G)C\Big)\tau_{3}\right)\Bigg)\rangle,\\
    I_{b0\uparrow,\downarrow} &=\langle\frac{T_{1}}{|d_{R}|^{2}}\Bigg((1+T_{1}^{2})\text{Tr}\left((G+G^{\dagger})(C+C^{\dagger})\right)\nonumber\\&+2T_{1}\text{Tr}\left((1+GG^{\dagger})(1+C^{\dagger}C)\right)\Bigg)\rangle,\\
    I_{b2}&=\langle\frac{T_{1}}{|d_{R}|^{2}}\Bigg(2T_{1}\text{Tr}\left(1+G\tau_{3}G^{\dagger}\tau_{3}+C\tau_{3}C^{\dagger}\tau_{3}+CG\tau_{3}G^{\dagger}C^{\dagger}\right)\nonumber\\&+(1+T_{1}^{2})\text{Tr}\left(CG+G^{\dagger}C^{\dagger}+G\tau_{3}C^{\dagger}\tau_{3}+C\tau_{3}G^{\dagger}\tau_{3}\right)\Bigg)\rangle\\
    I_{b3}&=\langle\frac{T_{1}}{|d_{R}|^{2}}\Bigg((1+T_{1}^{2})\text{Tr}\left(\Big(CG+CG^{\dagger}+GC^{\dagger}+G^{\dagger}C^{\dagger}\Big)\tau_{3}\right)\nonumber\\&+2T_{1}\left(\Big(GG^{\dagger}+CC^{\dagger}+CGG^{\dagger}C^{\dagger}\Big)\tau_{3}\right)\Bigg)\rangle\\
    I_{bs2}&=\langle\frac{T_{1}}{|d_{R}|^{2}}\Bigg((1+T_{1}^{2})\text{Tr}\left(GC+C^{\dagger}G^{\dagger}+G\tau_{3}C^{\dagger}\tau_{3}+C\tau_{3}G^{\dagger}\tau_{3}\right)\nonumber\\&+2T_{1}\text{Tr}\left(1+C\tau_{3}C^{\dagger}\tau_{3}+G\tau_{3}G^{\dagger}\tau_{3}+GC\tau_{3}C^{\dagger}G^{\dagger}\tau_{3}\right)\Bigg)\rangle\\
    d_{R}& = 1+T_{1}^{2}+T_{1}\text{Tr}(CG+GC).
\end{align}
In these expressions $f_{S\uparrow,\downarrow}$ is the distribution function in the superconductor and $C$ and $G$ are the retarded quantities studied in the main text. The subscripts for spin $\uparrow,\downarrow$ and for Keldysh space $R,A,K$ have been suppressed in the quantities $C$ and $G$ for clarity of notation. The notation $\langle\rangle$ is used for angular averaging and $T_{1} = \frac{T}{2-T+2\sqrt{1-T}}$, where $T$ is the transparency of the junction.
From this it follows that
\begin{align}
    \text{Tr}(I^{K\uparrow,\downarrow}) & = I_{b2\uparrow,\downarrow}f_{L\uparrow,\downarrow}(0)-I_{bs2\uparrow,\downarrow}f_{S\uparrow,\downarrow}(0)+I_{b3\uparrow,\downarrow}f_{T\uparrow,\downarrow}(0).
\end{align}
Following \cite{tanaka2021phys} it holds that
\begin{align}
    I_{b1} & := I_{b1\uparrow} = -I_{b1\downarrow}\\
    I_{bs} & := I_{bs\uparrow} = -I_{bs\downarrow}\\
    I_{b0} & := I_{b0\uparrow} = I_{b0\downarrow}.
\end{align}
Therefore, using that $f_{S\uparrow} = f_{S\downarrow} = \tanh{\frac{E}{kT}}=:\frac{1}{2}f_{S}$
\begin{align}
    X_{2} := \text{Tr}((I^{K\uparrow}+I^{K\downarrow})\tau_{3}) & = I_{b1}f_{L}(0) + I_{b0}f_{T}(0),\label{eq:Ik3}\\
    X_{1} := \text{Tr}((I^{K\uparrow}-I^{K\downarrow})) & = I_{b2}f_{L}(0) + I_{b3}f_{T}(0).\label{eq:Ik0}
\end{align}
Equations \ref{eq:fL} to \ref{eq:fTBC}, together with expressions \ref{eq:Ik3} and \ref{eq:Ik0} form a linear system of equations with nonconstant coefficients for $f_{L}$ and $f_{T}$, independent from $f_{L\uparrow}+f_{L\downarrow}$ and $f_{T\uparrow}-f_{T\downarrow}$. It reduces to the equations shown in \cite{tanaka2021phys} for $I_{b1} = C = 0$. In the coming subsection the equations as found in this section will be solved.\\
\subsection{Solution}
From equation \ref{eq:fL} it follows that
\begin{align}
    \nabla f_{L} = \frac{1}{D_{L}}\frac{1}{\gamma_{B}L}X_{1}-\frac{\Tilde{C}}{D_{L}}\nabla f_{T}.
\end{align}
Substituting this in equation \ref{eq:fT} it follows that
\begin{equation}
    (D_{T}+\frac{\Tilde{C}^{2}}{D_{L}})\nabla f_{T} = \frac{1}{\gamma_{B}L}(X_{2}+\frac{\Tilde{C}}{D_{L}}X_{1}).
\end{equation}
Similarly, it follows that
\begin{equation}
    (D_{L}+\frac{\Tilde{C}^{2}}{D_{T}})\nabla f_{L} = \frac{1}{\gamma_{B}L}(X_{1}-\frac{\Tilde{C}}{D_{T}}X_{2}).
\end{equation}
These equations can be solved to give
\begin{align}
    f_{T}(0) &= f_{T0}-\frac{1}{\gamma_{B}L}\int_{0}^{L}\frac{1}{D_{T}+\frac{\Tilde{C}^{2}}{D_{L}}}dx X_{2}-\frac{1}{\gamma_{B}L}\int_{0}^{L}\frac{\Tilde{C}}{D_{L}D_{T}+\Tilde{C}^{2}}dx X_{1}.\label{eq:fTintegrals}\\
    f_{L}(0) &= -\frac{1}{\gamma_{B}L}\int_{0}^{L}\frac{1}{D_{L}+\frac{\Tilde{C}^{2}}{D_{T}}}dx X_{1} + \frac{1}{\gamma_{B}L}\int_{0}^{L}\frac{\Tilde{C}}{D_{L}D_{T}+\Tilde{C}^{2}}dx X_{2}.\label{eq:fLintegrals}
\end{align}
By filling in expressions \ref{eq:Ik0} and \ref{eq:Ik3} into equation \ref{eq:fLintegrals} an equation relating $f_{L}(0)$ and $f_{T}(0)$ is found:
\begin{align}
    f_{L}(0) =& X_{3}f_{T}(0),\label{eq:fLX3}\\
    X_{3} =& \left(1+\frac{I_{b2}}{\gamma_{B}L}\int_{0}^{L}\frac{1}{D_{L}+\frac{\Tilde{C}^{2}}{D_{T}}}dx-\frac{I_{b1}}{\gamma_{B}L}\int_{0}^{L}\frac{\Tilde{C}}{D_{L}D_{T}+\Tilde{C}^{2}}dx\right)^{-1}\nonumber\\&\left(-\frac{I_{b3}}{\gamma_{B}L}\int_{0}^{L}\frac{1}{D_{L}+\frac{\Tilde{C}^{2}}{D_{T}}}dx+\frac{I_{b0}}{\gamma_{B}L}\int_{0}^{L}\frac{\Tilde{C}}{D_{L}D_{T}+\Tilde{C}^{2}}dx\right).
\end{align}
An important observation about $X_{3}$ is that in the limit $I_{b1} = 0$ and $\Tilde{C} = 0$, it holds that $X_{3} = 0$. This means that in this limit $f_{L}(0) = 0$, that is, $f_{L\uparrow}(0) = f_{L\downarrow}(0)$ as expected.\\
By filling in equations \ref{eq:Ik0}, \ref{eq:Ik3} and \ref{eq:fLX3} into \ref{eq:fTintegrals}, an expression for $f_{T}(0)$ is found:
\begin{align}
    f_{T}(0) =& f_{T0}-\frac{I_{b0}}{L\gamma_{B}}X_{4}f_{T}(0),\label{eq:fT0fT(0)}\\
    X_{4} =& \int_{0}^{L}\frac{1}{D_{T}+\frac{\Tilde{C}^{2}}{D_{L}}}dx+\frac{I_{b3}}{I_{b0}}\int_{0}^{L}\frac{\Tilde{C}}{D_{L}D_{T}+C^{2}}dx+X_{3}\left(\frac{I_{b1}}{I_{b0}}\int_{0}^{L}\frac{1}{D_{T}+\frac{\Tilde{C}^{2}}{D_{L}}}dx+\frac{I_{b2}}{I_{b0}}\int_{0}^{L}\frac{\Tilde{C}}{D_{L}D_{T}+\Tilde{C}^{2}}dx\right).
\end{align}
Note that in the limit $I_{b1} = \Tilde{C} = 0$ only the first term in the expression of $X_{4}$ is nonzero.\\
Using expressions \ref{eq:fLX3} and \ref{eq:fT0fT(0)} the expression for the current through the junction becomes
\begin{align}
    I_{e} &= \frac{2}{e}\int_{0}^{\infty}\frac{1}{R_{b}}\left(I_{b0}f_{T}(0)+I_{b1}f_{L}(0)\right)dE\nonumber\\
    & = \frac{2}{e}\int_{0}^{\infty}\frac{f_{T0}}{R_{b}}\left(I_{b0}\frac{1}{1+\frac{I_{b0}}{L\gamma_{B}}X_{4}}+I_{b1}\frac{X_{3}}{1+\frac{I_{b0}}{L\gamma_{B}}X_{4}}\right)dE\nonumber\\
    & = \frac{2}{e}\int_{0}^{\infty}f_{T0}\left(\frac{1}{\frac{R_{b}}{I_{b0}}+\frac{R_{d}}{L}X_{4}}+\frac{X_{3}}{\frac{R_{b}}{I_{b1}}+\frac{R_{d}}{L}\frac{I_{b0}}{Ib_{1}}X_{4}}\right)dE.
\end{align}
From this it follows that the conductance at zero temperature is given by
\begin{align}
    R &= \frac{1}{2}\left(\frac{1}{\frac{R_{b}}{I_{b0}}+\frac{R_{d}}{L}X_{4}}+\frac{X_{3}}{\frac{R_{b}}{I_{b1}}+\frac{R_{d}}{L}\frac{I_{b0}}{I_{b1}}X_{4}}\right)^{-1}.\\
    & = \frac{1}{2}(1+X_{3}\frac{I_{b1}}{I_{b0}})^{-1}\left(\frac{R_{b}}{I_{b0}}+\frac{R_{d}}{L}X_{4}\right) \label{eq:Rexpression}
\end{align}
There are a few important remarks to be made about expression \ref{eq:Rexpression}:
\begin{itemize}
\item As remarked, in the limit $I_{b1} = \Tilde{C} =0$, it holds that $X_{3} =0 $ and $X_{4} =\int_{0}^{L}\frac{1}{D_{T}}dx$, which means that in this limit the resistance $R_{\text{lim}}$ is given by
\begin{equation}
    R_{\text{lim}} = \frac{1}{2}\left(\frac{R_{b}}{I_{b0}}+\frac{R_{d}}{L}\int_{0}^{L}\frac{1}{D_{T}}dx\right),
\end{equation}
which is the expression found in \cite{tanaka2021phys}.
\item If $C$ and $I_{b1}$ flip sign, that is, if $\text{Im}(\chi)$ flips sign, then $X_{3}$ flips sign, whereas $X_{4}$ remains the same. This means that also the resistance remains the same upon a sign flipping of $\text{Im}(\chi)$.
\end{itemize}
\subsection{Exploration of limit}
In the introduction of this section it was claimed that for pure s-wave or pure p-wave superconductors the expression as found in \cite{tanaka2021phys} was valid. Thereafter, it was shown that the new equations reduce to the equations in \cite{tanaka2021phys} if $I_{b1} = \Tilde{C} = 0$. To proof the statement, it will now be shown that in pure s-wave and pure p-wave superconductors the relation $I_{b1} = \Tilde{C} = 0$ is satisfied. First it will be shown that $\Tilde{C} = 0$ holds for pure s-wave and pure p-wave superconductors, thereafter it is shown that $I_{b1} = 0$ for such superconductors.\\
Recall that that $C$ is proportional to $\sinh{2\text{Im}(\chi)}$. Thus, $\Tilde{C}$ vanishes if $\text{Im}(\chi) = 0$. The quantity $\text{Im}(\chi)$ is zero if
\begin{equation}|f_{+}+f_{-}+(g_{-}f_{+}-g_{+}f_{-})| = |f_{+}+f_{-}-(g_{-}f_{+}-g_{+}f_{-})|\label{eq:Imchivoorwaarde}.\end{equation} Now, for s-wave superconductors $f_{+} = f_{-}$ and $g_{+} = g_{-}$, so  equation \ref{eq:Imchivoorwaarde} reduces to $|f_{+}+f_{-}| = |f_{+}+f_{-}|$, which is trivially satisfied. On the other hand, for p-wave superconductors, $f_{+} = -f_{-}$ and $g_{+} = g_{-}$, which means that the expression reduces to $|g_{-}f_{+}-g_{+}f_{-}| = |-(g_{-}f_{+}-g_{+}f_{-})|$, which is also trivially satisfied. Thus, for pure s-wave and pure p-wave superconductors, $\Tilde{C} = 0$ is satisfied.\\
Now, recall the expression for $I_{b1}$ from equation \ref{eq:Ib1expressionupdown}:
\begin{align}
    I_{b1}&=\langle\frac{T_{1}}{|d_{R}|^{2}}\text{Tr}\left(\Big((C+C^{\dagger})G+G^{\dagger}(C+C^{\dagger})\Big)\tau_{3}\right)+\text{Tr}\left(\Big(1+C^{\dagger}C+G^{\dagger}(1+C^{\dagger}C)G\Big)\tau_{3}\right)\rangle,\nonumber
\end{align}
where the subscript $\uparrow$ in $C$ and $G$ has been suppressed for clarity of presentation. Now, by virtue of the absence of a spectral supercurrent, the $\chi$-equation for the retarded part gives $\langle\frac{T_{1}}{|d_{R}|^{2}}\text{Tr}(CG\tau_{3})\rangle = 0$.
Using conjugation, this immediately implies that
$\langle\frac{T_{1}}{|d_{R}|^{2}}\text{Tr}(G^{\dagger}C^{\dagger}\tau_{3})\rangle = 0$. This holds for both s-wave and p-wave superconductors. Moreover $\text{Tr}(\tau_{3}) = 0$. These terms are thus zero for both s-wave and p-wave superconductors and any mixture of the two.\\
For the other terms the s-wave and p-wave cases need to be considered separately. 
First consider the following general types of matrices and their rules: Let $\psi\in\mathbb{C}$ be fixed. Then
\begin{itemize}
    \item Matrices of Type I are matrices of the form $A = \begin{bmatrix}a&be^{i\psi}\\-be^{-i\psi}&-a\end{bmatrix}$.
    \item Matrices of Type II are matrices of the form $B = \begin{bmatrix}c&de^{i\psi}\\de^{-i\psi}&-c\end{bmatrix}$.
    \item If $A$ is of Type I and $B$ of Type II, then $A^{\dagger}$ and $B^{\dagger}$ are as well of Type I and Type II respectively.
    \item For matrix $A_{1}$ and $A_{2}$ of type I, $A_{1}^{\dagger}A_{2} = \begin{bmatrix}a_{1}^{*}a_{2}+b_{1}^{*}b_{2}&(a_{1}^{*}b_{2}+b_{1}^{*}a_{2})e^{i\psi}\\(a_{1}^{*}b_{2}+b_{1}^{*}a_{2})e^{-i\psi}&a_{1}^{*}a_{2}+b_{1}^{*}b_{2}\end{bmatrix}$, and hence $\text{Tr}(A_{1}^{\dagger}A_{2}\tau_{3}) = 0$.
    \item If $B_{1}$ and $B_{2}$ are of type II, then $B_{1}^{\dagger}B_{2} = \begin{bmatrix}c_{1}^{*}c_{2}+d_{1}^{*}d_{2}&(c_{1}^{*}d_{2}-d_{1}^{*}c_{2})e^{i\psi}\\-(c_{1}^{*}d_{2}-d_{1}^{*}c_{2})e^{-i\psi}&c_{1}^{*}c_{2}+d_{1}^{*}d_{2}\end{bmatrix}$, and hence $\text{Tr}(B_{1}^{\dagger}B_{2}\tau_{3}) = 0$.
\end{itemize}
First consider s-wave superconductors. Then $H_{-} = 0$, so $C = H_{+}^{-1} = \begin{bmatrix}g_{+}&f_{+}e^{i\psi}\\-f_{+}e^{-i\psi}&-g_{+}\end{bmatrix}$. 
Moreover, recall that for s-wave superconductors the $\chi$-equation gives $\chi = \psi\in\mathbb{R}$, so we may write
\begin{equation}
    G = \begin{bmatrix}\cosh{\theta}&\sinh{\theta}e^{i\psi}\\-\sinh{\theta}e^{-i\psi}&-\cosh{\theta}\end{bmatrix}.
\end{equation}
Thus, both $C$ and $G$ are of type I. By the third and fourth bullet points it now follows that for all modes
\begin{equation}
\text{Tr}(C^{\dagger}G\tau_{3}) = \text{Tr}(G^{\dagger}C\tau_{3}) = \text{Tr}(G^{\dagger}G\tau_{3}) = \text{Tr}(C^{\dagger}C\tau_{3}) = 0.\end{equation} This only leaves $\text{Tr}(G^{\dagger}C^{\dagger}CG\tau_{3}) = \text{Tr}((CG)^{\dagger}CG\tau_{3})$. But according to the fourth rule above, $CG$ is of the form $K = \begin{bmatrix}k&le^{i\psi}\\le^{-i\psi}&k\end{bmatrix}$. It can be calculated that for such matrices $K^{\dagger}K = \begin{bmatrix}|k|^{2}+|l|^{2}&(k^{*}l+kl^{*})e^{i\psi}\\(k^{*}l+kl^{*})e^{-i\psi}&|k|^{2}+|l|^{2}\end{bmatrix}$. Thus, for all modes,
\begin{equation}
    \text{Tr}((CG)^{\dagger}CG\tau_{3})  = 0.   
\end{equation}
This shows that $I_{b1} = 0$ for s-wave superconductors.\\
For p-wave superconductors, $H_{+}^{-1} = \frac{1}{g_{+}}\tau_{3}$ and $H_{-} = \begin{bmatrix}0&f_{+}e^{i\psi}&-f_{+}e^{-i\psi}\end{bmatrix}$, and therefore
$C = \frac{1}{g_{+}}\begin{bmatrix}1&-f_{+}e^{i\psi}\\-f_{+}e^{-i\psi}&-1\end{bmatrix}$, so $C$ is of type II. Moreover, the $\chi$-equation implies now that $\chi = \psi-\frac{\pi}{2}$, therefore $G = \begin{bmatrix}\cosh{\theta}&-i\sinh{\theta}e^{i\psi}\\-i\sinh{\theta}e^{-i\psi}&-\cosh{\theta}\end{bmatrix}$, so also $G$ is of type II.
By the third and fifth bullet points above it now follows that for all modes
\begin{equation}
\text{Tr}(C^{\dagger}G\tau_{3}) = \text{Tr}(G^{\dagger}C\tau_{3}) = \text{Tr}(G^{\dagger}G\tau_{3}) = \text{Tr}(C^{\dagger}C\tau_{3}) = 0.\end{equation} This only leaves $\text{Tr}(G^{\dagger}C^{\dagger}CG\tau_{3}) = \text{Tr}((CG)^{\dagger}CG\tau_{3})$. 
But according to the fifth rule above, this is of the form $M = \begin{bmatrix}m&ne^{i\psi}\\-ne^{-i\psi}&m\end{bmatrix}$. It can be calculated that for such matrices $M^{\dagger}M = \begin{bmatrix}|m|^{2}+|n|^{2}&(m^{*}n-nl^{*})e^{i\psi}\\(-m^{*}n+mn^{*})e^{-i\psi}&|m|^{2}+|n|^{2}\end{bmatrix}$. Thus, for all modes,
\begin{equation}
    \text{Tr}((CG)^{\dagger}CG\tau_{3})  = 0.   
\end{equation}
This shows that $I_{b1} = 0$ for p-wave superconductors as well.\\
Thus, for both pure s-wave and pure p-wave superconductors, $I_{b1} = \Tilde{C} = 0$. This shows that the formula in \cite{tanaka2021phys} is indeed valid for pure s-wave and pure p-wave superconductors. Deviations from the formula in \cite{tanaka2021phys} occur if both $f_{+}+f_{-}$ and $g_{+}f_{-}-g_{-}f_{+}$ are nonzero.
\subsection{Numerical comparison to standard method}
\begin{figure}
    \centering
    \subcaptionbox{$r = 0$
  }[0.45\linewidth]
  {\hspace*{-2em}\includegraphics[width =8.4cm]{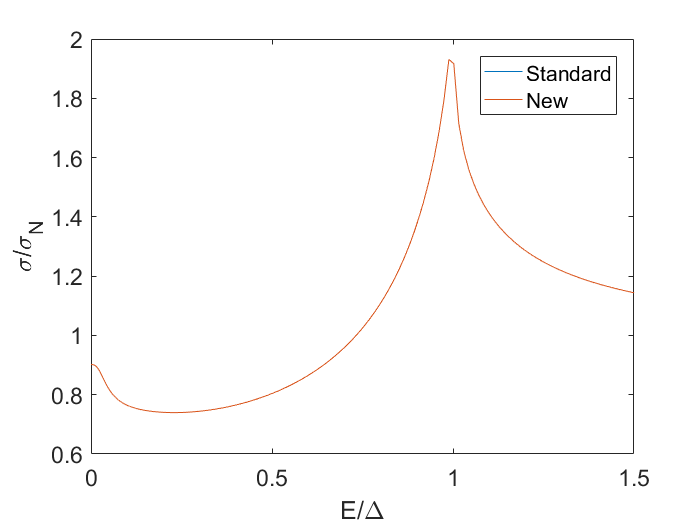}}
\hfill
\subcaptionbox{$r = 0.2$
  }[0.45\linewidth]
  {\hspace*{-2em}\includegraphics[width =8.4cm]{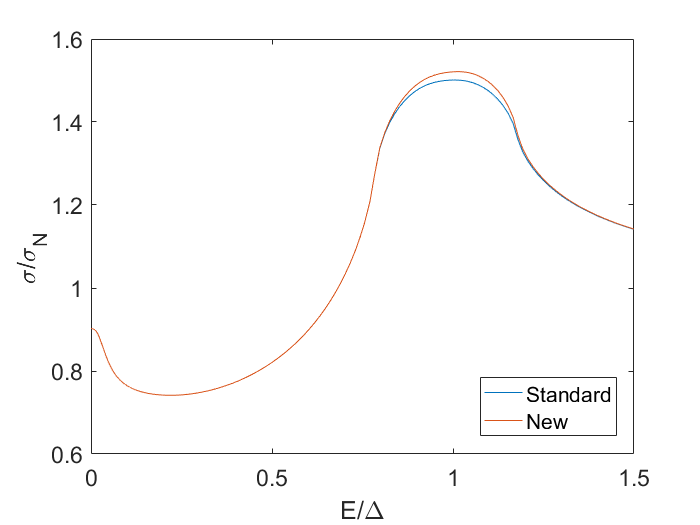}}
\hfill
\subcaptionbox{$r = 0.9$
  }[0.45\linewidth]
  {\hspace*{-2em}\includegraphics[width =8.4cm]{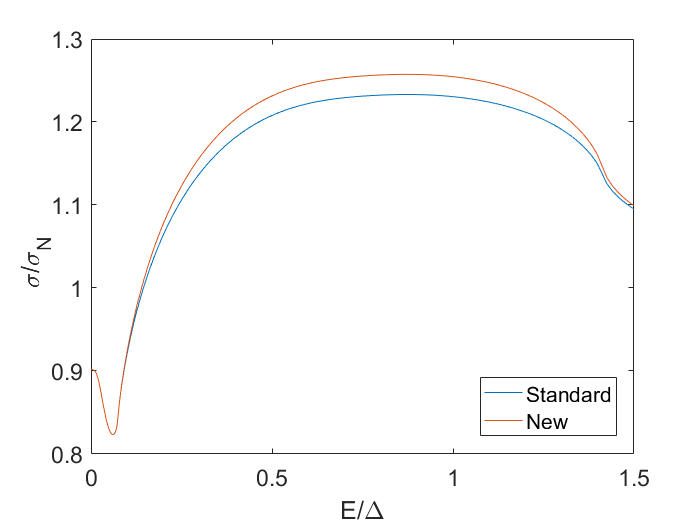}}
\hfill
\subcaptionbox{$r = 2$
  }[0.45\linewidth]
  {\hspace*{-2em}\includegraphics[width =8.4cm]{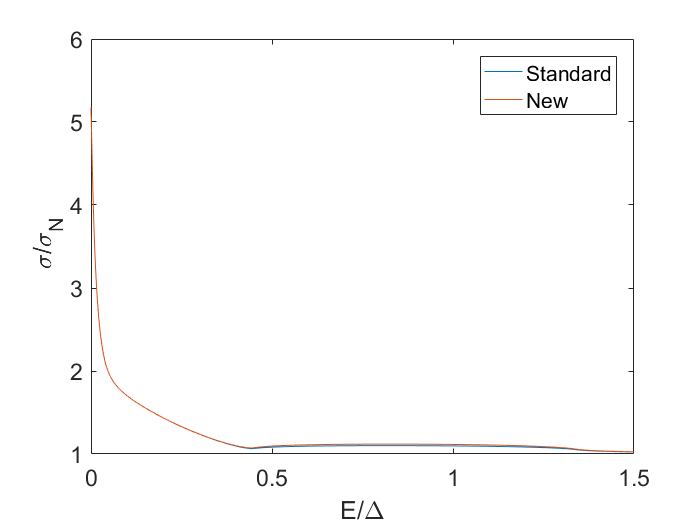}}
\hfill
\subcaptionbox{$r = 5$
  }[0.45\linewidth]
  {\hspace*{-2em}\includegraphics[width =8.4cm]{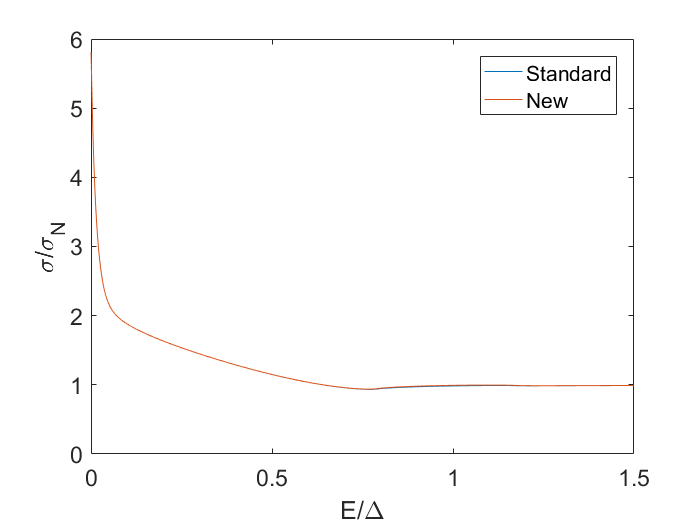}}
\hfill
\subcaptionbox{$r = \infty$
  }[0.45\linewidth]
  {\hspace*{-2em}\includegraphics[width =8.4cm]{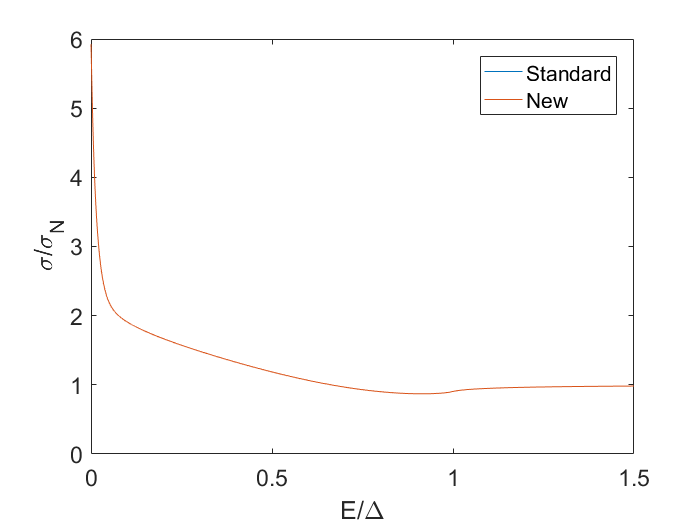}}
    \caption{Comparison of the conductance $\sigma$ as calculated by the new conductance formula and the conductance as calculated using the standard formula for the s + p-wave junction. As expected, the two methods give the same results for pure s-wave and pure p-wave junctions. However, if both singlet and triplet components appear, the new conductance formula shows an increase of the conductance that is mainly apparent around $E = \Delta$ and more visible if the singlet and triplet components are more equal.}
    \label{fig:Comparer}
\end{figure}

\begin{figure}
    \centering
    \subcaptionbox{$r = 0$
  }[0.45\linewidth]
  {\hspace*{-2em}\includegraphics[width =8.4cm]{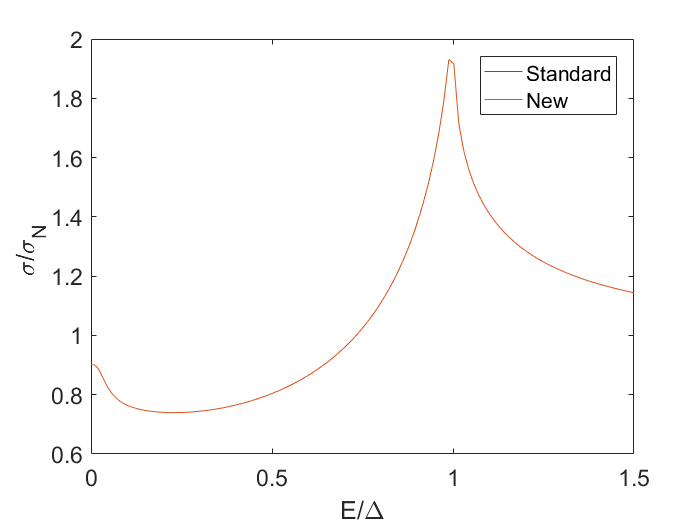}}
\hfill
\subcaptionbox{$r = 0.2$
  }[0.45\linewidth]
  {\hspace*{-2em}\includegraphics[width =8.4cm]{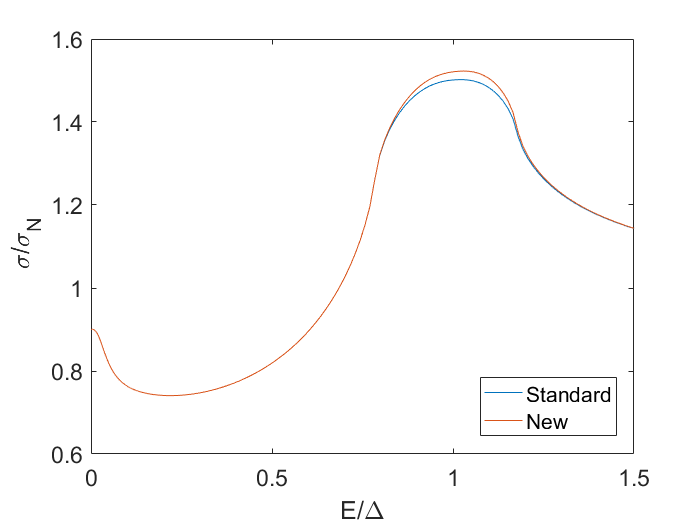}}
\hfill
\subcaptionbox{$r = 0.9$
  }[0.45\linewidth]
  {\hspace*{-2em}\includegraphics[width =8.4cm]{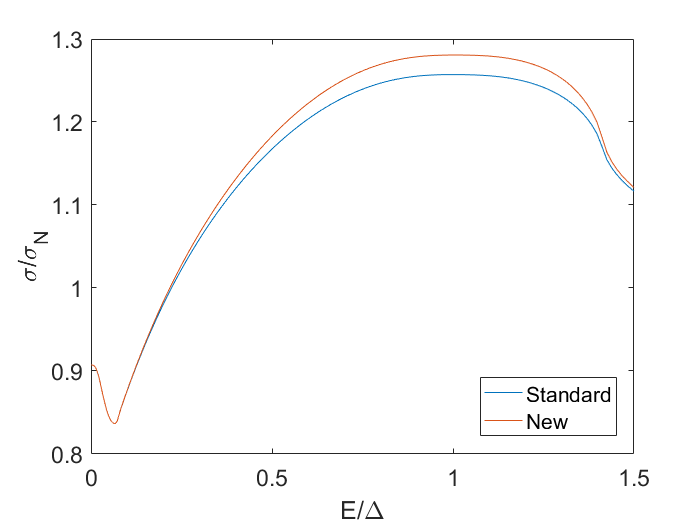}}
\hfill
\subcaptionbox{$r = 2$
  }[0.45\linewidth]
  {\hspace*{-2em}\includegraphics[width =8.4cm]{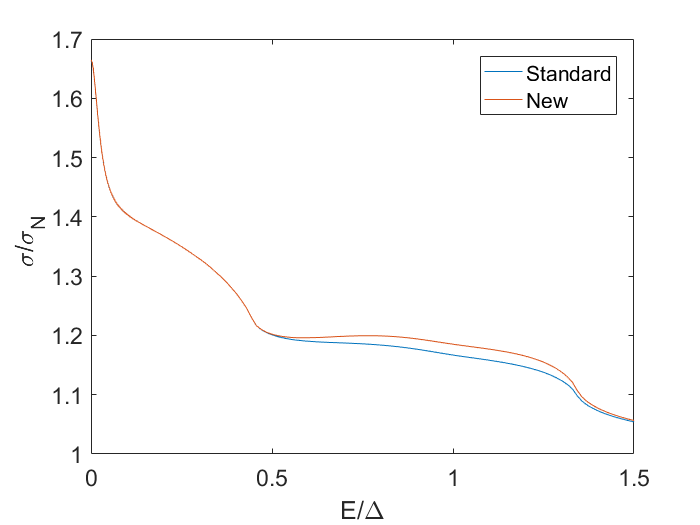}}
\hfill
\subcaptionbox{$r = 5$
  }[0.45\linewidth]
  {\hspace*{-2em}\includegraphics[width =8.4cm]{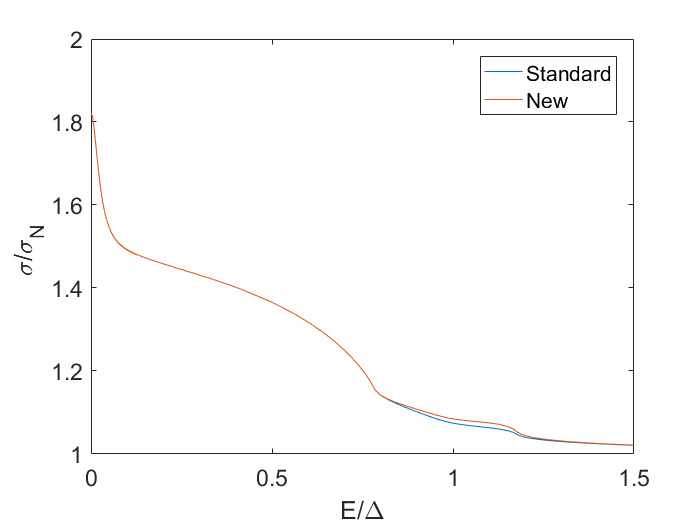}}
\hfill
\subcaptionbox{$r = \infty$
  }[0.45\linewidth]
  {\hspace*{-2em}\includegraphics[width =8.4cm]{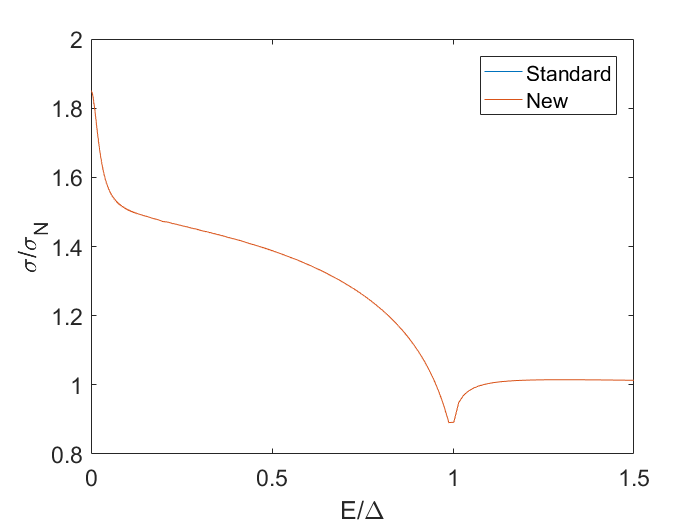}}
    \caption{Comparison of the conductance $\sigma$ as calculated by the new conductance formula and the conductance as calculated using the standard formula for the s + chiral p-wave junction. As expected, the two methods give the same results for pure s-wave and pure chiral p-wave junctions. However, if both singlet and triplet components appear, the new conductance formula shows an increase of the conductance that is mainly apparent around $E = \Delta$ and more visible if the singlet and triplet components are more equal.}
    \label{fig:CompareChiralr}
\end{figure}

Both the new expression for the resistance \ref{eq:Rexpression} and the expression from \cite{tanaka2021phys}, which will be called hte standard method were implemented in MATLAB. The results for the two-dimensional s + p and s + chiral p junctions are shown in figures \ref{fig:Comparer} and \ref{fig:CompareChiralr} respectively. As expected, in the case of pure type superconductors, the two expressions yield the same result. However, if $\Delta_{s}$ and $\Delta_{p}$ are both nonzero, the new formula predicts an increase of the conductance that is mainly apparent for energies close to $\Delta_{0}$. The effect of the new terms is similar for p-wave junctions as for chiral p-wave junctions, though the difference for $r>1$ is more apparent for chiral p-wave junctions, because the conductance around $E = \Delta_{0}$ is significantly larger for chiral p-wave junctions.

\subsection{Conductance calculations for the s + helical p-wave superconductor junctions}\label{sec:Helical conductance}
As discussed in the main body, also the Keldysh part of the Green's function satisfies $YGY = G$. However, if the superconductor pair potential is a mixture of s-wave and helical p-wave potentials, the expressions for the boundary terms need to be adjusted. In this subsection the derivation of the boundary term in the presence of a helical p-wave superconductor is shown.\\
As shown in \cite{tanaka2021phys}, the current through the interface $I$ is given in Nambu-Keldysh space by
\begin{equation}
    \frac{I}{2T_{1}} = \begin{bmatrix}d_{R}&T_{1}(N_{1}+N_{2})\\0&d_{A}\end{bmatrix}^{-1}\begin{bmatrix}[C^{R},G^{R}]&N_{1}-N_{2}\\0&[C^{A},G^{A}]\end{bmatrix},
\end{equation}
where the notation $[\cdot,\cdot]$ is used to denote the commutator of two matrices, and
\begin{align}
    d_{R} & = (1+T_{1}^{2})\mathbf{1}+T_{1}(G^{R}C^{R}+C^{R}G^{R}),\\
    d_{A} & = \tau_{3}d_{R}^{\dagger}\tau_{3},\\
    N_{1} & = (CG)^{K} = C^{R}G^{K}+C^{K}G^{A},\\
    N_{2} & = (GC)^{K} = G^{R}C^{K}+G^{K}C^{A},
\end{align}
where $C_{R,A,K}$ and $G_{R,A,K}$ are respectively the retarded part, the advanced part and the Keldysh part of the matrices $C$ and $G$, and $\mathbf{1}$ the four by four identity matrix. As helical p-wave superconductors are considered, $C^{R}$ and $G^{R}$ are four by four matrices. This makes an important difference. 
If $C^{R}$ and $G^{R}$ are two by two matrices, as for the $\text{p}_{\text{x}}$-wave and chiral p-wave superconductors, the anticommutator is proportional to the identity matrix. Thus, for mixture of s-wave superconductors and $\text{p}_{\text{x}}$-wave and chiral p-wave superconductors the matrix $d_{R}$ is proportional to the identity matrix, and the expression in \cite{tanaka2021phys} can be used. On the other hand, the anticommutator of four by four matrices need not be proportional to the identity matrix, therefore $d_{R}$ is in general not proportional to the identity matrix if helical superconductors are considered. In this case, $I$ can be expressed as
\begin{equation}
    \frac{I}{2T_{1}} = \begin{bmatrix}d_{R}^{-1}&-T_{1}d_{R}^{-1}(N_{1}+N_{2})d_{A}\\0&d_{A}^{-1}\end{bmatrix}\begin{bmatrix}[C^{R},G^{R}]&N_{1}-N_{2}\\0&[C^{A},G^{A}]\end{bmatrix}.
\end{equation}
From this it can be read that the Keldysh component of the current reads
\begin{align}
    \frac{I^{K}}{2T_{1}} &= d_{R}^{-1}(N_{1}-N_{2})-T_{1}d_{R}^{-1}(N_{1}+N_{2})d_{A}^{-1}[C^{A},G^{A}]\\
    &=d_{R}^{-1}\left(N_{1}(1-T_{1}(C^{A}G^{A}-G^{A}C^{A}))-N_{2}(1+T_{1}(C^{A}G^{A}-G^{A}C^{A}))\right)d_{A}^{-1},\label{eq:IdRdA}
\end{align}
where it was used that $d_{A}$ commutes with $C^{A}G^{A}$ and $G^{A}C^{A}$, since the identity matrix commutes with any matrix, and the anti commutator of two matrices $A,B$ clearly commutes with both $AB$ and $BA$. Expression \ref{eq:IdRdA} is of the form $d_{R}^{-1}\left(...\right)d_{A}^{-1}$. The part between $\left(...\right)$ was shown in \cite{tanaka2021phys} to equal
\begin{align}
    &(1+T_{1}^{2})(C^{R}G^{K}+C^{K}G^{A}-G^{R}C^{K}-G^{K}C^{A})\label{eq:ConductanceHelical1}\\
    &+T_{1}((C^{R}G^{K}+C^{K}G^{A})G^{A}C^{A}-(G^{R}C^{K}+G^{K}C^{A})C^{A}G^{A})\label{eq:ConductanceHelical2}
\end{align}
Now, the quantities $C^{K}$ and $G^{K}$ can be parametrised as
\begin{align}
    C^{K} &= f_{S}\left(C^{R}-C^{A}\right),\\
    G^{K} &= f_{l}\left(G^{R}-G^{A}\right)+f_{L}\left(G^{R}\omega_{1}-\omega_{1}G^{A}\right)+f_{t}\left(G^{R}\omega_{2}-\omega_{2}G^{A}\right)+f_{T}\left(G^{R}\omega_{3}-\omega_{3}G^{A}\right),\label{eq:GKparametrised}\\
    \omega_{1} &= \text{diag}(1,-1,-1,1),\\
    \omega_{2} &= \text{diag}(1,-1,1,-1),\\
    \omega_{3} &= \text{diag}(1,1,-1,-1)
\end{align}
This corresponds to the definitions $f_{L} = f_{L\uparrow}-f_{L\downarrow}$, $f_{l} = f_{L\uparrow}+f_{L\downarrow}$, $f_{T} = f_{T\uparrow}+f_{T\downarrow}$, $f_{t} = f_{T\uparrow}-f_{T\downarrow}$. It should be noted that under the assumption $f_{L\uparrow} = f_{L\downarrow}$ and $f_{T\uparrow} = f_{T\downarrow}$ only the first and last term in equation \ref{eq:GKparametrised} survive. Here the general derivation in which all terms in equation \ref{eq:GKparametrised} are allowed to be nonzero will be given.\\
Using that $G^{A} = -\omega_{3}(G^{R})^{\dagger}\omega_{3}$, $C^{A} = -\omega_{3}(C^{R})^{\dagger}\omega_{3}$, $\omega_{1}\omega_{3} = \omega_{2} = \omega_{3}\omega_{1}$ and $\omega_{2}\omega_{3} = \omega_{1} = \omega_{3}\omega_{2}$, expression \ref{eq:ConductanceHelical1} can be written as 
\begin{align}
    &-f_{S}\left(C^{R}\omega_{3}(G^{R})^{\dagger}\omega_{3}+\omega_{3}(C^{R})^{\dagger}(G^{R})^{\dagger}\omega_{3}+G^{R}C^{R}+G^{R}\omega_{3}(C^{R})^{\dagger}\omega_{3}\right)\nonumber\\
    &+f_{l}\left(C^{R}G^{R}+C^{R}\omega_{3}(G^{R})^{\dagger}\omega_{3}+G^{R}\omega_{3}(C^{R})^{\dagger}\omega_{3}+\omega_{3}(G^{R})^{\dagger}(C^{R})^{\dagger}\omega_{3}\right)\nonumber\\
    &+f_{L}\left(C^{R}G^{R}\omega_{1}+C^{R}\omega_{2}(G^{R})^{\dagger}\omega_{3}+G^{R}\omega_{2}(C^{R})^{\dagger}\omega_{3}+(G^{R})^{\dagger}\omega_{2}(C^{R})^{\dagger}\omega_{3}\right)\nonumber\\
    &+f_{t}\left(C^{R}G^{R}\omega_{2}+C^{R}\omega_{1}(G^{R})^{\dagger}\omega_{3}+G^{R}\omega_{1}(C^{R})^{\dagger}\omega_{3}+(G^{R})^{\dagger}\omega_{1}(C^{R})^{\dagger}\omega_{3}\right)\nonumber\\
    &+f_{T}\left(C^{R}(G^{R}+(G^{R})^{\dagger})+(G^{R}+(G^{R})^{\dagger})(C^{R})^{\dagger}\right)\omega_{3}\label{eq:2T1terms}
\end{align}
Meanwhile, expression \ref{eq:ConductanceHelical2} can be written as
\begin{align}
    &-f_{S}\left(\mathbf{1}+C^{R}\omega_{3}(C^{R})^{\dagger}\omega_{3}+G^{R}\omega_{3}(G^{R})^{\dagger}\omega_{3}+G^{R}C^{R}\omega_{3}(C^{R})^{\dagger}(G^{R})^{\dagger}\omega_{3}\right)\nonumber\\
    &+f_{l}\left(\mathbf{1}+G^{R}\omega_{3}(G^{R})^{\dagger}\omega_{3}+C^{R}\omega_{3}(C^{R})^{\dagger}\omega_{3}+C^{R}G^{R}\omega_{3}(G^{R})^{\dagger}(C^{R})^{\dagger}\omega_{3}\right)\nonumber\\
    &+f_{L}\left(\omega_{1}+G^{R}\omega_{2}(G^{R})^{\dagger}\omega_{3}+C^{R}\omega_{2}(C^{R})^{\dagger}\omega_{3}+C^{R}G^{R}\omega_{2}(G^{R})^{\dagger}(C^{R})^{\dagger}\omega_{3}\right)\nonumber\\
    &+f_{t}\left(\omega_{2}+G^{R}\omega_{1}(G^{R})^{\dagger}\omega_{3}+C^{R}\omega_{1}(C^{R})^{\dagger}\omega_{3}+C^{R}G^{R}\omega_{1}(G^{R})^{\dagger}(C^{R})^{\dagger}\omega_{3}\right)\nonumber\\
    &+f_{T}\left(\omega_{3}+G^{R}(G^{R})^{\dagger}\omega_{3}+C^{R}(C^{R})^{\dagger}\omega_{3}+C^{R}G^{R}(G^{R})^{\dagger}(C^{R})^{\dagger}\omega_{3}\right)\label{eq:1+T1squareterms}
\end{align}
In the derivations for the $\text{p}_{\text{x}}$-wave and chiral p-wave cases, the expressions $\text{Tr}((I^{K\uparrow}+I^{K\downarrow})\tau_{3})$ and $\text{Tr}(I^{K\uparrow}-I^{K\downarrow})$ were needed for the calculation of conductance, whereas $\text{Tr}((I^{K\uparrow}-I^{K\downarrow})\tau_{3})$ and $\text{Tr}(I^{K\uparrow}+I^{K\downarrow})$ were used for the calculation of $f_{l}$ and $f_{t}$. In the four by four matrix representations these quantities are given by:
\begin{align}
    \text{Tr}((I^{K\uparrow}+I^{K\downarrow})\tau_{3}) &= \text{Tr}(I^{K}\omega_{3})\\
    \text{Tr}(I^{K\uparrow}-I^{K\downarrow}) &= \text{Tr}(I^{K}\omega_{1})\\
    \text{Tr}((I^{K\uparrow}-I^{K\downarrow})\tau_{3}) &=\text{Tr}(I^{K}\omega_{2})\\
    \text{Tr}(I^{K\uparrow}+I^{K\downarrow}) &= \text{Tr}(I^{K})
\end{align}
Now, first it will be shown that, as for the mixtures between an s-wave potential and $\text{p}_{\text{x}}$-wave or chiral p-wave potentials, the former two only involve $f_{L}$ and $f_{T}$, and the latter only involve $f_{l}$ and $f_{t}$, that is, the equations for $f_{L,T}$ are decoupled from the equations for $f_{l,t}$, as for the s + $\text{p}_{\text{x}}$-wave and s + chiral p-wave junctions.\\
As discussed in section \ref{sec:Qsymm}, the Green's function $G^{R}$ and the quantity $C^{R}$ satisfy $QGQ = G$ and $QCQ = C$ for 
\begin{equation}
    Q = \begin{bmatrix}0&0&1&0\\0&0&0&1\\-1&0&0&0\\0&-1&0&0\end{bmatrix}.
\end{equation}

Moreover,
\begin{align}
    Q\omega_{1}Q &= \omega_{1}\\
    Q\omega_{3}Q &= \omega_{3}, 
\end{align}
but
\begin{align}
    Q\omega_{2}Q &= -\omega_{2}\\
    Q^{2} &=-\mathbf{1}. \label{eq:Qsquare} 
\end{align}
These Q-symmetries will be used to show that $\text{Tr}I^{K}$ and $\text{Tr}I^{K}\omega_{3}$ can be simplified. It will be shown that whereas there are terms proportional to the five quantities $f_{S},f_{l},f_{L},f_{t},f_{T}$ in expressions \ref{eq:2T1terms} and \ref{eq:1+T1squareterms} for $I^{K}$, only the terms involving $f_{L}$ and $f_{T}$ give a nonzero contribution for $\text{Tr}I^{K}$ and $\text{Tr}I^{K}\omega_{3}$.\\
First, from equation \ref{eq:Qsquare} an important relation can be deduced.
Suppose $V = \prod_{i = 1}^{n}V_{i}$ is a matrix that is the product of any $n$ matrices $V_{i}$, where $n\in\mathbb{N}$ is arbitrary., then
\begin{equation}\label{eq:QVQ}
    QVQ = Q (\prod_{i = 1}^{n}V_{i})Q = (-1)^{n-1}\prod_{i = 1}^{n}(QV_{i}Q).
\end{equation}
According to equation \ref{eq:Qsquare} if $W$ is the product of $p$ matrices $M_{i}$, i = 1...p of type M, that is, such that $QM_{i}Q = M_{i}$,  matrices and q matrices $N_{j}$, j = 1...q, of type N, that is, such that $QN_{j}Q = -N_{j}$, then the following equation holds:
\begin{align}
QWQ = (-1)^{(p+q-1)}\prod_{i = 1}^{p}QM_{i}Q\prod_{j = 1}^{q} QN_{j}Q = (-1)^{(p+q-1)}\prod_{i = 1}^{p}(-M_{i})\prod_{j = 1}^{q} N_{j} = 
(-1)^{q}(-1)^{(p+q-1)}W = (-1)^{p-1}W,
\end{align}
where the factor $(-1)^{q}$ appears from the $q$ matrices of type $N$ and the factor $-(1)^{p+q-1}$ from the fact that $p+q$ matrices are multiplied, i.e. $n = p+q$. From this it follows that
\begin{equation}
    \text{Tr}(W) = \text{Tr}(-Q^{2}W) = -\text{Tr}(QWQ) = (-1)^{p}\text{Tr}(W).
\end{equation}
If $p$ is an odd integer this implies $\text{Tr}(W) = 0$. Now, as seen from expressions \ref{eq:2T1terms} and \ref{eq:1+T1squareterms}, the terms involving $f_{S}, f_{l}, f_{t}$ all contain an even number of Type M matrices, whereas all terms involving $f_{L}$ and $f_{T}$ involve an odd number of Type M matrices. Therefore, the terms containing $f_{L}$ and $f_{T}$ will drop out of $\text{Tr}(I^{K})$ and $\text{Tr}(I^{K}\omega_{2})$, whereas the terms involving $f_{S}, f_{l}, f_{t}$ drop out of $\text{Tr}(I^{K}\omega_{1})$ and $\text{Tr}(I^{K}\omega_{3})$, exactly as when a $\text{p}_{\text{x}}$-wave potential or a chiral p-wave potential is involved.\\
Using this
\begin{align}
\text{Tr}(I^{K}\omega_{3}) & = I_{b1}f_{L}(0) + I_{b0}f_{T}(0),\label{eq:Ik3helical}\\
\text{Tr}(I^{K}\omega_{1}) & = I_{b2}f_{L}(0) + I_{b3}f_{T}(0).\label{eq:Ik0helical}
\end{align}
Using that $\omega_{3}d_{A}^{-1}\omega_{3} = (d_{R}^{\dagger})^{-1}$, $\omega_{3}d_{A}^{-1}\omega_{1} = (d_{R}^{\dagger})^{-1}\omega_{2}$ and it follows that
\begin{align}
    I_{b0} &= (1+T_{1}^{2})\text{Tr}\left(d_{R}^{-1}\Bigg(C^{R}(G^{R}+(G^{R})^{\dagger})+(G^{R}+(G^{R})^{\dagger})(C^{R})^{\dagger}\Bigg)(d_{R}^{\dagger})^{-1}\right)\nonumber\\&+2T_{1}\text{Tr}\left(d_{R}^{-1}\Bigg(\mathbf{1}+G^{R}(G^{R})^{\dagger}+C^{R}(C^{R})^{\dagger}+C^{R}G^{R}(G^{R})^{\dagger}(C^{R})^{\dagger}\Bigg)(d_{R}^{\dagger})^{-1}\right).\label{eq:Ib0Helical}\\
    I_{b1}&= (1+T_{1}^{2})\text{Tr}\left(d_{R}^{-1}\Bigg(C^{R}G^{R}\omega_{2}+C^{R}\omega_{2}(G^{R})^{\dagger}+G^{R}\omega_{2}(C^{R})^{\dagger}+(G^{R})^{\dagger}\omega_{2}(C^{R})^{\dagger}\Bigg)(d_{R}^{\dagger})^{-1}\right)\nonumber\\&+2T_{1}\text{Tr}\left(d_{R}^{-1}\Bigg(\omega_{2}+G^{R}\omega_{2}(G^{R})^{\dagger}+C^{R}\omega_{2}(C^{R})^{\dagger}+C^{R}G^{R}\omega_{1}(G^{R})^{\dagger}(C^{R})^{\dagger}\Bigg)(d_{R}^{\dagger})^{-1}\right)\\
    I_{b2} &= (1+T_{1}^{2})\text{Tr}\left(d_{R}^{-1}\Bigg(C^{R}G^{R}\omega_{2}+C^{R}\omega_{2}(G^{R})^{\dagger}+G^{R}\omega_{2}(C^{R})^{\dagger}+(G^{R})^{\dagger}\omega_{2}(C^{R})^{\dagger}\Bigg)(d_{R}^{\dagger})^{-1}\omega_{2}\right)\nonumber\\&+2T_{1}\left(d_{R}^{-1}\Bigg(\omega_{2}+G^{R}\omega_{2}(G^{R})^{\dagger}+C^{R}\omega_{2}(C^{R})^{\dagger}+C^{R}G^{R}\omega_{2}(G^{R})^{\dagger}(C^{R})^{\dagger}\Bigg)(d_{R}^{\dagger})^{-1}\omega_{2}\right)\\
    I_{b3} &= (1+T_{1}^{2})\left(d_{R}^{-1}\Bigg(C^{R}(G^{R}+(G^{R})^{\dagger})+(G^{R}+(G^{R})^{\dagger})(C^{R})^{\dagger}\Bigg)(d_{R}^{\dagger})^{-1}\omega_{2}\right)\nonumber\\&+2T_{1}\left(d_{R}^{-1}\Bigg(\mathbf{1}+G^{R}(G^{R})^{\dagger}+C^{R}(C^{R})^{\dagger}+C^{R}G^{R}(G^{R})^{\dagger}(C^{R})^{\dagger}\Bigg)(d_{R}^{\dagger})^{-1}\omega_{2}\right)
\end{align}
Similar expressions can be found for $\text{Tr}(I^{K})$ and $\text{Tr}(I^{K}\omega_{2})$ that yield expressions only involving $f_{l}$ and $f_{t}$.
With the adaption of the expressions for the boundary currents the formulas discussed in the previous sections are still valid.
\\
It should be noted that under the assumption that $f_{L\uparrow} = f_{L\downarrow}$ and $f_{T\uparrow} = f_{T\downarrow}$, the expression for $\text{Tr}((I^{K\uparrow}+I^{K\downarrow})\tau_{3})$ becomes
\begin{align}
    \text{Tr}(I^{K}\omega_{2}) = I_{b0}f_{T},
\end{align}
with $I_{b0}$ given as above. This expression is very similar to the expression in \cite{tanaka2021phys}, with the division by the absolute square of the trace of $d_{R}$ now replaced by left multiplication by $d_{R}^{-1}$ and right multiplication by $(d_{R}^{\dagger})^{-1}$. In the case $d_{R}$ is proportional to the identity matrix the two expressions thus give the same result, as required. In that case the formula for the resistance given in \cite{tanaka2021phys} can be used with the replacement of $I_{b0}$ by equation \ref{eq:Ib0Helical}.
\section{Calculation of the spin conductance in the ballistic junction}
In \cite{tanaka2009theory}, the quantities $a_{\uparrow,\uparrow},a_{\downarrow,\uparrow},a_{\uparrow,\downarrow},a_{\downarrow,\downarrow},b_{\uparrow,\uparrow},b_{\downarrow,\uparrow},b_{\uparrow,\downarrow},b_{\downarrow,\downarrow}$ were introduced such that in the normal metal the spinor as introduced in section \ref{sec:Superconductivityandspin} is given by
\begin{align}
    \Psi_{N}e^{-ik_{y}y} = \begin{bmatrix}1\\0\\a_{\uparrow,\uparrow}\\a_{\uparrow,\downarrow}\end{bmatrix}e^{ikx}+\begin{bmatrix}b_{\uparrow,\uparrow}\\b_{\uparrow,\downarrow}\\0\\0\end{bmatrix}e^{-ikx}
\end{align}
or
\begin{align}
    \Psi_{N}e^{-ik_{y}y} = \begin{bmatrix}0\\1\\a_{\downarrow,\uparrow}\\a_{\downarrow,\downarrow}\end{bmatrix}e^{ikx}+\begin{bmatrix}b_{\downarrow,\uparrow}\\b_{\downarrow,\downarrow}\\0\\0\end{bmatrix}e^{-ikx},
\end{align}
where $k = \sqrt{2m\mu}$,
and in the superconducting material the spinor is given by
\begin{align}
    \Psi_{S}e^{-ik_{y}y} = c_{1}\begin{bmatrix}u_{1}\\-i\frac{u_{1}}{\alpha_{1}}\\i\frac{v_{1}}{\alpha_{1}}\\v_{1}\end{bmatrix}e^{iq_{1}^{+}x}+c_{2}\begin{bmatrix}v_{1}\\-i\frac{v_{1}}{\Tilde{\alpha}_{1}}\\i\frac{u_{1}}{\Tilde{\alpha}_{1}}\\u_{1}\end{bmatrix}e^{-iq_{1}^{-}x}+c_{3}\begin{bmatrix}u_{2}\\i\frac{u_{2}}{\alpha_{2}}\\i\gamma\frac{v_{2}}{\alpha_{2}}\\-\gamma v_{2}\end{bmatrix}e^{iq_{2}^{+}x}+c_{4}\begin{bmatrix}v_{2}\\i\frac{v_{2}}{\Tilde{\alpha}_{2}}\\i\gamma\frac{u_{2}}{\Tilde{\alpha}_{2}}\\-\gamma u_{2}\end{bmatrix}e^{-iq_{2}^{+}x},
\end{align}
where $\gamma = \text{sign}(\Delta_{p}-\Delta_{s})$, $u_{1,2} = \sqrt{\frac{1}{2}+\frac{1}{2}\frac{\sqrt{E^{2}-\Delta_{1,2}^{2}}}{E}}$,$v_{1,2} = \sqrt{\frac{1}{2}-\frac{1}{2}\frac{\sqrt{E^{2}-\Delta_{1,2}^{2}}}{E}}$, $\Delta_{1} = \Delta_{s}+\Delta_{p}$, $\Delta_{2} = |\Delta_{s}-\Delta_{p}|$, $\alpha_{1,2} = \cos{\phi}-i\sin{\phi}$, $\Tilde{\alpha}_{1,2} = -\cos{\phi}-i\sin{\phi}$, and
\begin{align}
    q_{1,2}^{\pm} &= k_{1,2}\cos{\phi}\pm\frac{1}{\cos{\phi}}\sqrt{\frac{E^{2}-\Delta_{1,2}^{2}}{\lambda^{2}}+2\frac{\mu}{m}},\\
    k_{1} & = -m\lambda+\sqrt{(m\lambda)^{2}+2m\mu}\\
    k_{2} & = m\lambda+\sqrt{(m\lambda)^{2}+2m\mu}
\end{align}
where $\mu$ is the Fermi level, $m$ is the mass of electrons, and $\lambda$ is a parameter controlling the strength of spin orbit coupling.\\
The superconducting and normal metal wavefunctions at $x = 0$ satisfy the relations $\Psi_{N} =\Psi_{s}$ and $\pdv{H_{N}}{k_{x}}\Psi_{N}-\pdv{H_{S}}{k_{x}}\Psi_{S} = iU\tau_{3}\Psi_{N}$, where $\tau_{3} = \text{diag}(1,1,-1,-1)$ and $U\in\mathbf{R}$, and $H_{S/N}$ are the Hamiltonians in the normal and superconducting state listed in \cite{tanaka2009theory}. The spin current is expressed as \cite{tanaka2009theory}:
\begin{align}
    f_{S}(\phi) = \left(|a_{\uparrow,\uparrow}|^{2}-|a_{\uparrow,\downarrow}|^{2}-|b_{\uparrow,\uparrow}|^{2}+|b_{\uparrow,\downarrow}|^{2}+|a_{\downarrow,\uparrow}|^{2}-|a_{\downarrow,\downarrow}|^{2}-|b_{\downarrow,\uparrow}|^{2}+|b_{\downarrow,\downarrow}|^{2}\right)\frac{\cos\phi}{2}.
\end{align}
Now, the $\sigma_{x}$ and $\sigma_{y}$ spin conductance can be found by using incoming particles respectively $\frac{1}{\sqrt{2}}(1,\pm1,0,0)^{T}$ (x-spin) and $\frac{1}{\sqrt{2}}(1,\pm i,0,0)^{T}$ (y-spin) on the normal side instead of $(1,0,0,0)^{T}$ and $(0,1,0,0)^{T}$. By linearity the solutions to the equations with these type of incoming particles are
\begin{align}
    \frac{1}{\sqrt{2}}\begin{bmatrix}1\\\pm1\\a_{\uparrow,\uparrow}\pm a_{\downarrow,\uparrow}\\a_{\uparrow,\downarrow}\pm a_{\downarrow,\downarrow}\end{bmatrix}e^{ikx}+\begin{bmatrix}b_{\uparrow,\uparrow}\pm b_{\downarrow,\uparrow}\\b_{\uparrow,\downarrow}\pm b_{\downarrow,\downarrow}\\0\\0\end{bmatrix}e^{-ikx},
\end{align}
for spin x aligned particles and
\begin{align}
    \frac{1}{\sqrt{2}}\begin{bmatrix}1\\\pm i\\a_{\uparrow,\uparrow}\pm ia_{\downarrow,\uparrow}\\a_{\uparrow,\downarrow}\pm ia_{\downarrow,\downarrow}\end{bmatrix}e^{ikx}+\begin{bmatrix}b_{\uparrow,\uparrow}\pm ib_{\downarrow,\uparrow}\\b_{\uparrow,\downarrow}\pm ib_{\downarrow,\downarrow}\\0\\0\end{bmatrix}e^{-ikx},
\end{align}
for spin y aligned particles. Now, whereas for the z-component the coefficients could be found by projection of the vectors $\begin{bmatrix}a_{\uparrow,\uparrow}\\a_{\uparrow,\downarrow}\end{bmatrix}$,$\begin{bmatrix}a_{\downarrow,\uparrow}\\a_{\downarrow,\downarrow}\end{bmatrix}$, $\begin{bmatrix}b_{\uparrow,\uparrow}\\b_{\uparrow,\downarrow}\end{bmatrix}$,$\begin{bmatrix}b_{\downarrow,\uparrow}\\b_{\downarrow,\downarrow}\end{bmatrix}$ on the eigenvectors of $\sigma_{z}$, for the x-component and the y-component of the spin conductance, projections of $\begin{bmatrix}a_{\uparrow,\uparrow}\pm a_{\downarrow,\uparrow}\\a_{\uparrow,\downarrow}\pm a_{\downarrow,\downarrow}\end{bmatrix}$ and $\begin{bmatrix}b_{\uparrow,\uparrow}\pm b_{\downarrow,\uparrow}\\b_{\uparrow,\downarrow}\pm b_{\downarrow,\downarrow}\end{bmatrix}$ on the eigenvectors of $\sigma_{x}$ and of $\begin{bmatrix}a_{\uparrow,\uparrow}\pm ia_{\downarrow,\uparrow}\\a_{\uparrow,\downarrow}\pm ia_{\downarrow,\downarrow}\end{bmatrix}$ and $\begin{bmatrix}b_{\uparrow,\uparrow}\pm ib_{\downarrow,\uparrow}\\b_{\uparrow,\downarrow}\pm ib_{\downarrow,\downarrow}\end{bmatrix}$ on the eigenvectors of $\sigma_{y}$ are needed.\\
This implies that the x-component of the spin current is given by
\begin{align}
    f_{Sx} &= \Bigg(|a_{\uparrow,\uparrow}+a_{\downarrow,\uparrow}+a_{\uparrow,\downarrow}+a_{\downarrow,\downarrow}|^{2}-|a_{\uparrow,\uparrow}+a_{\downarrow,\uparrow}-a_{\uparrow,\downarrow}-a_{\downarrow,\downarrow}|^{2}-|b_{\uparrow,\uparrow}+b_{\downarrow,\uparrow}+b_{\uparrow,\downarrow}+b_{\downarrow,\downarrow}|^{2}\nonumber\\&+|b_{\uparrow,\uparrow}+b_{\downarrow,\uparrow}-b_{\uparrow,\downarrow}-b_{\downarrow,\downarrow}|^{2}+|a_{\uparrow,\uparrow}-a_{\downarrow,\uparrow}+a_{\uparrow,\downarrow}-a_{\downarrow,\downarrow}|^{2}-|a_{\uparrow,\uparrow}-a_{\downarrow,\uparrow}-a_{\uparrow,\downarrow}+a_{\downarrow,\downarrow}|^{2}\nonumber\\&-|b_{\uparrow,\uparrow}-b_{\downarrow,\uparrow}+b_{\uparrow,\downarrow}-b_{\downarrow,\downarrow}|^{2}+|b_{\uparrow,\uparrow}-b_{\downarrow,\uparrow}-b_{\uparrow,\downarrow}+b_{\downarrow,\downarrow}|^{2}\Bigg)\frac{\cos\phi}{4}.
\end{align}
Similarly, the $y$-component is given by
\begin{align}
    f_{Sy} &= \Bigg(|a_{\uparrow,\uparrow}+ia_{\downarrow,\uparrow}-ia_{\uparrow,\downarrow}+a_{\downarrow,\downarrow}|^{2}-|a_{\uparrow,\uparrow}+ia_{\downarrow,\uparrow}+ia_{\uparrow,\downarrow}-a_{\downarrow,\downarrow}|^{2}-|b_{\uparrow,\uparrow}+ib_{\downarrow,\uparrow}-ib_{\uparrow,\downarrow}+b_{\downarrow,\downarrow}|^{2}\nonumber\\&+|b_{\uparrow,\uparrow}+ib_{\downarrow,\uparrow}+ib_{\uparrow,\downarrow}-b_{\downarrow,\downarrow}|^{2}+|a_{\uparrow,\uparrow}-ia_{\downarrow,\uparrow}-ia_{\uparrow,\downarrow}-a_{\downarrow,\downarrow}|^{2}-|a_{\uparrow,\uparrow}-ia_{\downarrow,\uparrow}+ia_{\uparrow,\downarrow}+a_{\downarrow,\downarrow}|^{2}\nonumber\\&-|b_{\uparrow,\uparrow}-ib_{\downarrow,\uparrow}-ib_{\uparrow,\downarrow}-b_{\downarrow,\downarrow}|^{2}+|b_{\uparrow,\uparrow}-ib_{\downarrow,\uparrow}+ib_{\uparrow,\downarrow}+b_{\downarrow,\downarrow}|^{2}\Bigg)\frac{\cos\phi}{4}.
\end{align}
The spin conductances normalised to the normal metal conductance $\sigma_{N}$ are now given by
\begin{align}
    \frac{1}{\sigma_{N}}(\sigma_{\text{up}}-\sigma_{\text{down}})_{x,y,z} = \int_{-\frac{\pi}{2}}^{\frac{\pi}{2}}f_{Sx,y,z}d\phi.
\end{align}
\section{Non-dominant component}\label{sec:nondominant}
In this section it will be discussed why the non-dominant component is absent at $E = 0$, and why it is prominent for larger energies, more specific, in the region $\Delta_{+}>E>>|\Delta_{-}|$.\\
\subsection{Low energies}\label{sec: Very low energy}
First consider the case $E\approx 0$. In that case
\begin{align}
    g_{\pm} &= \frac{E}{\sqrt{E^{2}-\Delta_{\pm}^{2}}} \approx -iE\frac{1}{|\Delta_{\pm}|}\\
    f_{\pm} & = \frac{\Delta_{\pm}}{\sqrt{E^{2}-\Delta_{\pm}^{2}}}\approx-i\frac{\Delta_{\pm}}{|\Delta_{\pm}|}.
\end{align}
In case $r<1$ this gives
\begin{align}
    g_{+}+g_{-}&\approx-iE(\frac{1}{|\Delta_{+}|}+\frac{1}{|\Delta_{-}|}) \\
    & = -iE((\frac{\sqrt{r^{2}+1}}{1+r})+(\frac{\sqrt{r^{2}+1}}{1-r}))=-2iE\frac{\sqrt{r^{2}+1}}{1-r^{2}}.\\
    g_{+}-g_{-}&\approx-iE(\frac{1}{|\Delta_{+}|}-\frac{1}{|\Delta_{-}|}) \\
    & = -iE((\frac{\sqrt{r^{2}+1}}{1+r})-(\frac{\sqrt{r^{2}+1}}{1-r}))=-2iE\frac{r\sqrt{r^{2}+1}}{1-r^{2}}.\\
    f_{+}+f_{-}&\approx-i-i = -2i\\
    f_{+}-f_{-} &\approx -i((1-(\frac{E}{\Delta_{+}})^{2})^{-\frac{1}{2}}-(1-(\frac{E}{\Delta_{-}})^{2})^{-\frac{1}{2}}) \approx -\frac{1}{2}iE^{2}(\frac{1}{\Delta_{+}^{2}}-\frac{1}{\Delta_{-}^{2}}).
\end{align}
On the other hand, if $r>1$, this gives
\begin{align}
    g_{+}-g_{-}&\approx-iE(\frac{1}{|\Delta_{+}|}+\frac{1}{|\Delta_{-}|}) \\
    & = -iE((\frac{\sqrt{r^{2}+1}}{1+r})+(\frac{\sqrt{r^{2}+1}}{1-r}))=-2iE\frac{\sqrt{r^{2}+1}}{1-r^{2}}.\\
    g_{+}+g_{-}&\approx-iE(\frac{1}{|\Delta_{+}|}-\frac{1}{|\Delta_{-}|}) \\
    & = -iE((\frac{\sqrt{r^{2}+1}}{1+r})-(\frac{\sqrt{r^{2}+1}}{1-r}))=-2iE\frac{r\sqrt{r^{2}+1}}{1-r^{2}}.\\
    f_{+}-f_{-}&\approx-i-i = -2i\\
    f_{+}+f_{-} &\approx -i((1-(\frac{E}{\Delta_{+}})^{2})^{-\frac{1}{2}}-(1-(\frac{E}{\Delta_{-}})^{2})^{-\frac{1}{2}}) \approx -\frac{1}{2}iE^{2}(\frac{1}{\Delta_{+}^{2}}-\frac{1}{\Delta_{-}^{2}}).
\end{align}
Thus, both if $r<1$ and $r>1$, $g_{+}\pm g_{-}$ are first order in energy, whereas $f_{+}\pm f_{-}$ can be either zeroth order or second order in energy.
Thus, for $r<1$,
\begin{align}
    H_{+} &= \begin{bmatrix}0&-i\\i&0\end{bmatrix}\\
    H_{-} &= -iE\frac{r\sqrt{r^{2}+1}}{1-r^{2}}\begin{bmatrix}1&0\\0&-1\end{bmatrix}\\
    C& = H_{+}^{-1}-H_{+}^{-1}H_{-}\approx \begin{bmatrix}0&-i\\i&0\end{bmatrix},
\end{align}
where only the leading order in $E$ is kept.
On the other hand, for $r>1$
\begin{align}
    H_{+}& = -iE\frac{r\sqrt{r^{2}+1}}{1-r^{2}}\begin{bmatrix}1&0\\0&-1\end{bmatrix}\\
    H_{-} &= \begin{bmatrix}0&-i\\i&0\end{bmatrix}\\
    C& = H_{+}^{-1}-H_{+}^{-1}H_{-}\approx \frac{1i}{E}\frac{1-r^{2}}{r\sqrt{r^{2}+1}}\begin{bmatrix}1&i\\i&-1\end{bmatrix}.
\end{align}
From the expression for $r<1$ it is clear that $C$ is constant in this parameter range, and hence the zero-energy solution is independent of $r$ for $r<1$. For $r>1$ this is not so clear, since $C$ has a prefactor that depends on $r$. Moreover, because only the leading order in $E$ is taken into account, $C$ does not square to the identity matrix. The lower order terms make sure that $C^{2} = \mathbf{1}$, but their contribution to the current vanishes as $E\xrightarrow{}0$ and therefore they are not presented here. Notice that the prefactor diverges as $E\xrightarrow{}0$. 
Therefore, $(CG+GC)$ will dominate the denominator, and at low energies the boundary condition reduces to 
\begin{equation}
    I = 2(CG+GC)^{-1}(CG-GC).
\end{equation}
Clearly, prefactors of $C$ will cancel out here, therefore also in the range $r>1$ the zero energy solution is independent of $r$. From this, it can be seen that the low energy $C$ only depends on $\frac{\Delta_{+}}{|\Delta_{+}|}$ and $\frac{\Delta_{-}}{|\Delta_{-}|}$. This can also be used in two-dimensions, where the above derivation shows that the low energy $C$ only depends on $\frac{\Delta_{+}(\phi)}{|\Delta_{+}(\phi)|}$ and $\frac{\Delta_{-}(\phi)}{|\Delta_{-}(\phi)|}$. Because these quantities are now angular dependent, the behaviour at low energy should be discussed separately in each case.
\subsection{Higher energies}
Now consider the case $\Delta_{+}>E>>|\Delta_{-}|$. It will be shown that in this case the singlet and triplet components are more equal in magnitude. To this end it will be shown that $|\text{Im}(\chi)|$ is large. 
In the case $\Delta_{+}>E>>|\Delta_{-}|$ we have by Taylor expansion that
\begin{align}
    g_{-} &=\frac{E}{\sqrt{E-\Delta_{-}^{2}}} = 1+\frac{1}{2}(\frac{\Delta_{-}}{E})^{2} + O\Big((\frac{\Delta_{-}}{E})^{4}\Big), \\\
    f_{-} & = \frac{\Delta_{-}}{\sqrt{E-\Delta_{-}^{2}}}  =  \frac{\Delta_{-}}{E} + \frac{1}{2}(\frac{\Delta_{-}}{E})^{3} + O\Big((\frac{\Delta_{-}}{E})^{5}\Big),\\
    g_{+} & = \frac{E}{\sqrt{E^{2}-\Delta_{+}^{2}}},\\
    f_{+} & = \frac{\Delta_{+}}{\sqrt{E^{2}-\Delta_{+}^{2}}},
\end{align}
where $g_{+}$ and $f_{+}$ differ substantially from 1 and 0 respectively because $E<\Delta_{+}$.
Now recall that the expression for the phase in the junction is given by:
\begin{equation*}
    \chi = \psi-\frac{1}{2i}\ln\left(\frac{f_{p}+f_{m}+g_{m}f_{p}-g_{p}f_{m}}{f_{p}+f_{m}-g_{m}f_{p}+g_{p}f_{m}}\right).
\end{equation*}
Focussing on the denominator of this expression, it follows that
\begin{equation}
    f_{p}+f_{m}-g_{m}f_{p}+g_{p}f_{m} = (1+g_{+})\frac{\Delta_{-}}{E}+O\Big((\frac{\Delta_{-}}{E})^{2}\Big).
\end{equation}
Meanwhile, for the numerator of this expression,
\begin{equation}
    f_{p}+f_{m}+g_{m}f_{p}-g_{p}f_{m} = 2f_{+} + O\Big(\frac{\Delta_{-}}{E}\Big).
\end{equation}
This shows that
\begin{equation}
    e^{2|\text{Im}(\chi)|} = |\frac{f_{p}+f_{m}+g_{m}f_{p}-g_{p}f_{m}}{f_{p}+f_{m}-g_{m}f_{p}+g_{p}f_{m}}| = \frac{E}{\Delta_{-}}\left(|\frac{2f_{+}}{1+g_{+}}|+O\Big(\frac{\Delta_{-}}{E}\Big)\right)\approx|\frac{2f_{+}}{1+g_{+}}|\frac{E}{|\Delta_{-}|}>>1,
\end{equation}
But then 
\begin{equation}
    |\frac{\cos^{2}{\chi}-\sin^{2}{\chi}}{\cos^{2}{\chi}}| = \frac{|\cos{\text{Re}(\chi)}|}{\cosh{2\text{Im}(\chi)}}\leq \frac{1}{\cosh{2\text{Im}(\chi)}}\approx |\frac{1+g_{+}}{2f_{+}}|\frac{|\Delta_{-}|}{E}<<1,
\end{equation}
which shows that $F_{1}$ and $F_{2}$, and hence the singlet and triplet components, have similar magnitude.

\end{document}